\documentclass[twocolumn]{aastex7}
\usepackage{booktabs} 
\usepackage{array} 
\usepackage{xcolor}
\usepackage{amsmath}  
\usepackage{placeins}
\usepackage{dblfloatfix}

\usepackage{caption}
\usepackage{subcaption}
\usepackage{hyperref}
\usepackage[T1]{fontenc}
\usepackage{enumitem}
\usepackage{graphicx}	
\usepackage{amsmath,amssymb}  
\newcommand{\msun}{\ensuremath{M_\odot}}  

\begin{document}

\title[From Galaxy to BH Mergers]{One Merge to Rule Them All: From Galaxy \\ Interactions to Black Hole Mergers Using Horizon-AGN}

\author[0000-0002-0000-0000]{Ecaterina Leonova} 
\affiliation{GRAPPA, Anton Pannekoek Institute for Astronomy and Institute of High-Energy Physics, University of Amsterdam, Science Park 904, NL-1098 XH Amsterdam, The Netherlands}
\affiliation{University of Amsterdam, Science Park 904, NL-1098 XH Amsterdam, The Netherlands}
\email{e.leonova@uva.nl}

\author[0000-0000-0000-0000]{Marta Volonteri}
\affiliation{Institut d’Astrophysique de Paris, Sorbonne Université, CNRS, UMR 7095, 98 bis bd Arago, 75014 Paris, France}
\email{}

\author[0000-0000-0000-0000]{Clotilde Laigle}
\affiliation{Institut d’Astrophysique de Paris, Sorbonne Université, CNRS, UMR 7095, 98 bis bd Arago, 75014 Paris, France}
\email{}

\author[0000-0000-0000-0000]{Samaya Nissanke}
\affiliation{GRAPPA, Anton Pannekoek Institute for Astronomy and Institute of High-Energy Physics, University of Amsterdam, Science Park 904, NL-1098 XH Amsterdam, The Netherlands}
\affiliation{Deutsches Zentrum f\"ur Astrophysik (DZA), Postplatz 1, 02826 G\"orlitz, Germany}
\affiliation{Institut f{\"u}r Physik und Astronomie, Universit{\"a}t Potsdam, Haus 28, Karl-Liebknecht-Str. 24/25, 14476 Potsdam, Germany}
\email{}

\author[0000-0000-0000-0000]{Pascal A. Oesch}
\affiliation{Department of Astronomy, University of Geneva, Chemin Pegasi 51, 1290 Versoix, Switzerland}
\affiliation{Cosmic Dawn Center (DAWN)}
\affiliation{Niels Bohr Institute, University of Copenhagen, Jagtvej 128, DK-2200 Copenhagen N, Denmark}
\email{}
\author[0000-0000-0000-0000]{Yohan Dubois}
\affiliation{Institut d’Astrophysique de Paris, Sorbonne Université, CNRS, UMR 7095, 98 bis bd Arago, 75014 Paris, France}
\email{}

\begin{abstract}
Galaxy mergers are fundamental drivers of galaxy evolution and black hole (BH) growth across cosmic time. We use the \textsc{Horizon-AGN} simulation to investigate the fraction of galaxy pairs, the merger fraction, and the galaxy merger rate over a wide range of stellar masses and redshifts. To identify physically connected pairs, we adapt the \textit{Matthews Correlation coefficient} (MCC) framework, optimizing thresholds in projected distance and redshift difference, and compare our selection to commonly used criteria in the literature. We then connect the derived galaxy merger rates to supermassive BH mergers, tracking the evolution from galaxy interactions to BH coalescences, thereby reconstructing the full merger history. We find that the galaxy pair fraction, merger fraction, characteristic timescale, and merger rate all evolve strongly with both stellar mass and redshift, with higher-mass galaxies and earlier galaxies showing elevated merger activity. BHs exhibit a similar evolutionary trend, with the volume-averaged BH merger rate peaking around cosmic noon ($z\sim2\mbox{--}3$). Our results demonstrate a close correspondence between galaxy and BH cosmic histories. This work provides a comprehensive, simulation-based framework for linking galaxy and BH merger populations, and offers refined selection criteria for future observational studies, for forecasts of gravitational wave detections with LISA, and interpretation of Pulsar Timing Array results.

\end{abstract}

\keywords{galaxies: evolution --- galaxies: interactions --- quasars: supermassive black holes --- gravitational waves --- methods: numerical}



\section{Introduction}

In the standard $\Lambda$CDM cosmological paradigm, structure formation proceeds hierarchically, with smaller systems forming first and subsequently merging to build more massive structures \citep{White1978, Davis1985}. Galaxy mergers are a key mechanism within this framework, playing a pivotal role in shaping galaxy properties over cosmic time. Mergers are thought to drive a range of transformative processes: the build-up of stellar mass, morphological transitions (e.g., from disks to spheroids), the triggering of intense starburst activity, and the fueling of central supermassive black holes (BHs), which can in turn activate powerful active galactic nuclei (AGN) \citep{Toomre1972, Hopkins2006, Kaviraj2014,Naab2014, RodriguezGomez2015}.

Theoretical studies and cosmological hydrodynamical simulations consistently highlight the importance of mergers in the assembly of massive galaxies and the co-evolution of galaxies and BHs \citep{DiMatteo2005, Springel2005, Dubois2014, Schaye2015, Hopkins2008, Volonteri2020}. Mergers can lead to central starbursts through gas inflows \citep{Mihos1996}, influence bulge formation and kinematic heating \citep{Bournaud2005, Tacchella2019}, and are predicted to leave observable imprints in galaxy structure and color distributions \citep{Snyder2017}.

Observationally, constraining the frequency and impact of mergers remains challenging, especially at high redshift ($z \gtrsim 2$). Two primary approaches are commonly used: (i) morphological classifications based on disturbed structures and asymmetries \citep{Conselice2003, Lotz2008a, Kartaltepe2015}, and (ii) the close-pair method, which identifies galaxy pairs within specified projected separations and relative velocities as proxies for imminent mergers \citep{Patton2000, Lin2008, Bundy2009, Mantha2018}. The latter is widely used because of its direct interpretability within simulations, and its independence from morphological features that may be difficult to resolve at early epochs due to surface brightness dimming and instrumental resolution limits \citep{Lotz2008a}.

Galaxy mergers are generally categorized into two types based on the stellar mass ratio of the progenitor galaxies. Major mergers typically involve mass ratios $\geq 1:4$ and are often associated with strong dynamical effects such as morphological transformation, quenching of star formation, and bulge formation \citep{Barnes1992, Kormendy2004, Hopkins2009}. Minor mergers, defined by mass ratios $< 1:4$, are more frequent and can drive gradual stellar mass growth, disk heating, and modest starburst or AGN activity, particularly when multiple minor mergers occur over short timescales \citep{Bournaud2005, Kaviraj2014}. While major mergers have long been considered the dominant transformative events, recent studies emphasize that the cumulative impact of minor mergers may be equally important in shaping galaxy populations across cosmic time \citep{Kaviraj2014, Martin2018}.

However, translating pair fractions into reliable merger rates requires correcting for the probability that observed pairs will actually merge and the time it takes for them to do so. These observability and conversion factors are complex functions of redshift, stellar mass, mass ratio, separation, and environment \citep{Kitzbichler2008, Lotz2010a, RodriguezGomez2015, Snyder2019}. Recent studies have attempted to calibrate these corrections using cosmological simulations. For example, \citet{Ventou2019} used the Illustris simulation~\citep{Vogelsberger2014} to derive pair-to-merger probabilities and timescales, which they applied to MUSE deep-field spectroscopic data to extend pair-based merger rates out to $z \sim 6$. Similarly, \citet{OLeary2021} employed the EMERGE empirical model to provide analytic fitting functions for observability timescales, emphasizing their strong dependence on redshift and galaxy properties.

Cosmological simulations are essential in this context, offering a detailed and time-resolved view of galaxy assembly, and enabling rigorous comparisons between theoretical merger histories and observable proxies. The \textsc{Horizon-AGN} simulation \citep{Dubois2014} is particularly well suited for this task. It spans a comoving volume of $(100\,h^{-1}\,\mathrm{Mpc})^3$ and incorporates sophisticated treatments of gas cooling, star formation, AGN feedback, and BH physics. The associated lightcone catalog \citep{Laigle2019, Kraljic2020} allows for direct mock observations, facilitating comparisons with surveys like COSMOS, CANDELS, and JWST/FRESCO \citep{Koekemoer2011, Casey2023, Oesch2023}.

In this work, we leverage the \textsc{Horizon-AGN} simulation and its lightcone to identify galaxy pairs across a wide stellar mass range ($10^{9}\mbox{--}10^{11.5}\,\msun$) and redshifts up to $z \sim 6$. We analyze the evolution of the galaxy merger rate as a function of stellar mass, mass ratio, redshift, and projected separation. We implement observationally motivated selection criteria to directly compare with empirical pair studies, and contrast the resulting merger rates with those derived from the true merger histories traced in the simulation trees. Our approach builds on recent efforts to link pair observability to underlying merger dynamics in a cosmologically representative and physically motivated context \citep{Snyder2017, Martin2018, Ventou2019, Pfister2020,OLeary2021}.

In addition to galaxy mergers, we extend our analysis to BH mergers. BHs, believed to reside in the centers of most massive galaxies, coalesce after galaxy mergers under suitable conditions, producing strong gravitational wave signals in the nHz to mHz regime \citep{Mayer2007, Dotti2012}. These events are key targets for current and upcoming gravitational wave observatories: Pulsar Timing Arrays (PTAs) \citep{Agazie2023,Reardon2023,Xu2023,EPTA2024}, and the space-based Laser Interferometer Space Antenna (LISA), which will be sensitive to BHs with masses $10^4\mbox{--}10^7\,\msun$ out to $z \sim 10$ \citep{AmaroSeoane2017}.

Simulations like \textsc{Horizon-AGN} offer a unique opportunity to study the coalescence of BHs in cosmological environments. They provide insight into the timescales between galaxy and BH mergers, the dependence of BH merger rates on host galaxy properties and cosmic time, and the environments conducive to gravitational wave emission \citep{Salcido2016, Kelley2017}. Moreover, they enable predictions of BH merger rates and distributions of mass, redshift, and sometimes spin‚ quantities critical for the planning and interpretation of LISA and PTA signals \citep{Volonteri2020, Berti2008, Bonetti2019}.

By combining mock observational pair-based techniques with merger tree analyses of galaxies and BHs, our study bridges the gap between what is observed and what is physically occurring in the evolving Universe. This  approach helps to reduce systematic uncertainties in observational merger rate estimates and enhances our understanding of the joint evolution of galaxies and BHs.

This paper is organized as follows. In Section~\ref{sec:methods}, we describe the \textsc{Horizon-AGN} simulation, lightcone construction, and pair selection criteria. In Section~\ref{sec:results}, we describe our methodology for identifying galaxy and black hole mergers and computing merger rates. We then discuss the implications of our findings for galaxy evolution and gravitational-wave astrophysics, and summarize our conclusions in Section~\ref{sec:conclusion}.

\section{Simulation and Methods}
\label{sec:methods}

\subsection{The Horizon-AGN Simulation}

The Horizon-AGN simulation \citep{Dubois2014} is a large-scale cosmological hydrodynamical simulation designed to model galaxy formation and evolution across cosmic time in a $\Lambda$CDM cosmology. The chosen cosmology is consistent with WMAP-7 data \citep{Komatsu11}, with total matter density $\Omega_{\rm m}=0.272$, dark energy density $\Omega_{\Lambda}=0.728$, baryon density $ \Omega_{\rm b}=0.045$, a Hubble constant of $H_{0}=70.4 \rm \ km\,s^{-1}\, Mpc^{-1}$, and an amplitude of the matter power spectrum and power-law index of the primordial power spectrum of $\sigma_8=0.81$ and $n_{\rm s}=0.967$ respectively.

It uses the adaptive mesh refinement code \textsc{RAMSES} \citep{Teyssier2002} to evolve a comoving box of $(100\,h^{-1}\,\mathrm{Mpc})^3$, resolving the most refined regions down to $\Delta x=1\,\rm kpc$ according to a pseudo-Lagrangian refinement criterion that locally adjusts the resolution to the initial mass resolution. A new maximum level of refinement is triggered in the simulation each time the scale factor doubles.  The mass resolution is $8\times10^7\,\mathrm{M}_\odot$ for dark matter particles and $2\times10^6\,\mathrm{M}_\odot$ for stellar particles.

Gas cooling is implemented down to $10^4$\,K using metal-dependent cooling curves \citep{Sutherland1993}, and heating from a uniform ultraviolet background is included following \citet{Haardt1996}, beginning at redshift $z_\mathrm{reion} = 10$. Star formation is modeled using a Schmidt law with an efficiency $\varepsilon_\star = 0.02$ in cells where the hydrogen number density exceeds $n_0 = 0.1\,\mathrm{H\,cm^{-3}}$ \citep{Rasera2006, Dubois2008}. Stellar feedback includes energy injection from Type Ia and II supernovae and stellar winds, assuming a Salpeter initial mass function \citep{Salpeter1955}.

BHs are seeded in dense, high-dispersion gas regions with an exclusion radius of 50\,ckpc to prevent multiple seeds in the same galaxy. Accretion follows a boosted Bondi-Hoyle-Lyttleton prescription \citep{Booth2009}, and is capped at the Eddington limit with a radiative efficiency of $\varepsilon_{\rm r}=0.1$. BH feedback operates in two modes: thermal feedback above 1\,\% of the Eddington luminosity, and mechanical bipolar outflows below that threshold, injecting kinetic energy into the surrounding gas \citep{Dubois2012, Dubois2013}. 

A subgrid model for dynamical friction prevents artificial BH wandering due to resolution limits, incorporating a drag force consistent with the gas wake formalism of \citet{Ostriker1999} and calibrated following \citet{Chapon2013}. BHs are merged when they are separated by $\leqslant 4 \Delta x$, corresponding to 4\,kpc. Horizon-AGN has been shown to reproduce a broad range of galaxy and BH properties \citep{Dubois2016, Volonteri2016, Kaviraj2017}, making it a robust tool for interpreting observational data and constructing realistic mock catalogues.

\subsection{Galaxy Selection from the Lightcone and Merger Tree}
\label{sec:galaxyselection}

A lightcone was built on the fly from the {\sc Horizon-AGN} simulated box. The final lightcone is made of about 22000 portions of concentric shells of angular opening 1\,deg$^{2}$ extracted at each coarse time step. Galaxies have been extracted from the stellar particles in the lightcone using {\sc AdaptaHOP} \citep{Aubert2004}, with a density threshold of 178 times the average matter density of the Universe at that redshift. 

Galaxies were extracted independently in the lightcone and in the snapshot boxes at each redshift (with the same parametrisation of {\sc AdaptaHOP}). A merger tree of galaxies in the box was produced with {\sc TREEMAKER} \citep{tweed2009}. In order to connect each lightcone galaxy to their merger tree buit from the box, we matched galaxies in the lightcone and in the box based on their positions \citep[see e.g.~Figure~1 of][]{Pfister2020}. 

\subsection{Black Hole-Galaxy Matching and Black Hole Merger Tracking}
\label{sec:bhmatching}

We summarize here the main information about the approach. We invite the reader to refer to \cite{Volonteri2016} and \cite{Volonteri2020} for a detailed description. Since in Horizon-AGN BHs are not anchored in the center of galaxies, BHs have to be assigned to their host galaxies. We assign a BH to a halo+galaxy structure if the BH is within 10\,\% of the virial radius of the dark matter halo and also within twice the effective radius of the most massive galaxy within that halo. If multiple BHs meet 
the criteria, the most massive BH is defined as the central BH, and removed from the list.  We start from main halos and proceed hierarchically through sub-halos. Once all central BHs are assigned, we repeat the process to assign off-center BHs to galaxies. 

Two BHs are merged in the simulation when their separation falls below $4 \Delta x$ ($\sim 4 \,\rm kpc$). This can be flagged when a BH disappears between two timesteps; the companion BH is located by searching for the BH that was closest to the vanished one. Such pair is dubbed ``numerical merger''. We take care to remove from this list BH mergers that occur far from the center, as it is unlikely such BH would pair and merge if the real cross section of BHs were taken into account. In reality BHs merge at separations several orders of magnitude smaller than 4\,kpc, therefore we calculate in post-processing timescales for sub-resolution dynamical friction, hardening by interactions with stars and a circumbinary disc, and emission of gravitational waves. BH pairs that merge by $z=0$ after adding these timescales are called ``delayed mergers''. In Horizon-AGN the ID of BHs is conserved, and in a merger the ID of the most massive BH is used. This guarantees a unique descendant assignment at all times.

\subsection{Galaxy Pair Selection}
\label{sec:pairselection}

\begin{figure*}[tb]
    \centering
        \begin{minipage}{0.85\textwidth}
        \centering
        \vspace{-5cm}
        \includegraphics[width=\textwidth]{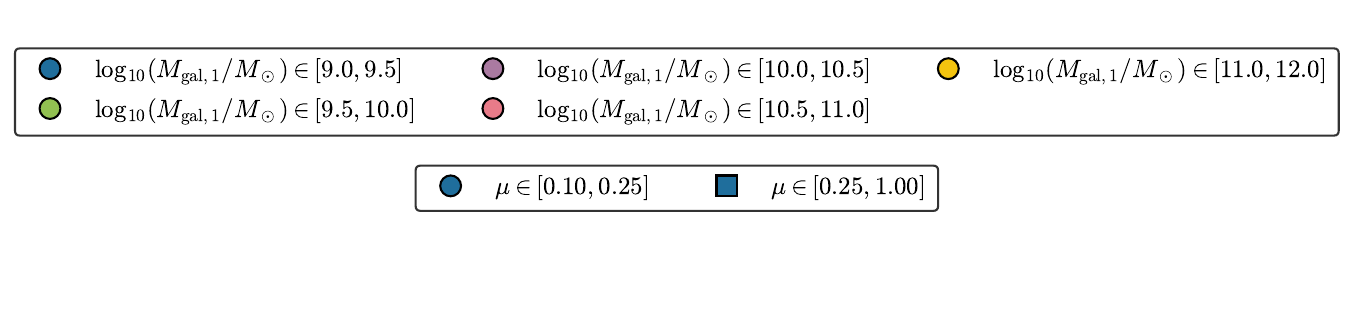}
    \end{minipage}

    \vspace{-2cm}
    \includegraphics[width=\textwidth]{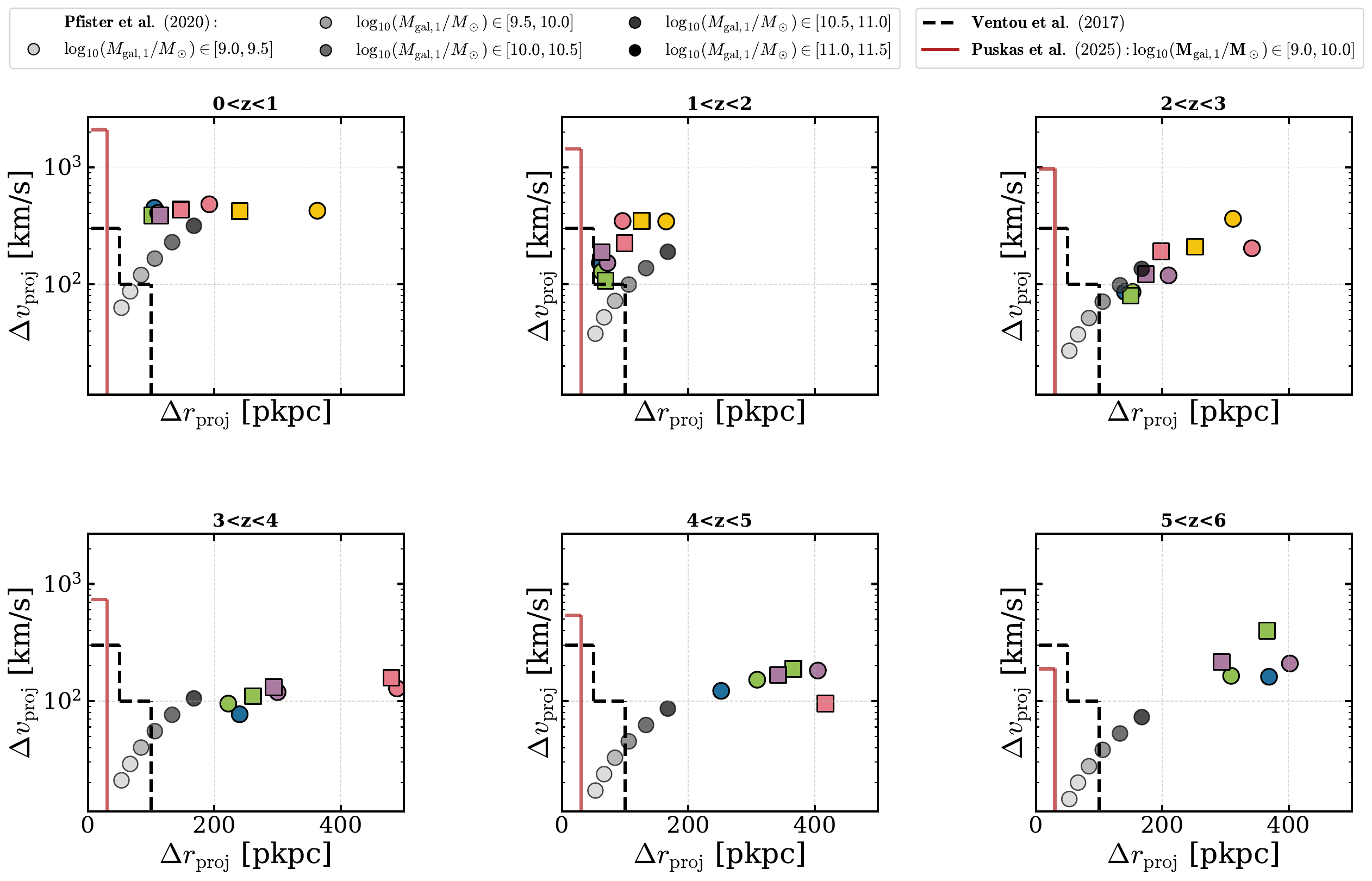} 
    \caption{
    Limits in projected separation $\Delta r_{\rm proj}$ and projected velocity difference $\Delta v_{\rm proj}$ for different redshift bins. Each panel corresponds to one redshift interval, as indicated in the titles. Coloured markers show our MCC derived from the Horizon-AGN lightcone, with circles representing major mergers ($\mu \in [0.25,1]$) and squares representing minor mergers ($\mu \in [0.1,0.25]$) for different stellar mass bins. Red lines indicate redshift limit criteria from \citet{Puskas2025}, grey circles show mass and redshift limits from \citet{Pfister2020}, and dashed black lines show limits from \citet{Ventou2019}. More massive galaxies require larger projected separation and velocity thresholds to capture all physically associated pairs. Our MCC-optimized criteria generally extend to wider separations than previous studies, reflecting the broader range of physically motivated merger conditions captured here.
    }
    \label{dvdrlim}
\end{figure*}

To identify close galaxy pairs in the lightcone catalogue, we apply a set of selection criteria designed to mimic observational strategies. Our work is therefore complementary to extracting the merger rate directly from the merger tree \citep{Kaviraj2015}. We begin by selecting galaxies with stellar masses above $10^9\,M_\odot$, from which galaxy pairs are identified based on redshift proximity, projected separation, and stellar mass ratio. To avoid duplicate counting, each galaxy is allowed to appear in at most one pair, with the nearest companion in projected distance retained when multiple candidates exist.

We adopt a redshift-dependent threshold on the redshift difference $\Delta z = |z_1 - z_2|$:
\[
\Delta z < 0.1 ,
\]
to account for computational limitations. 

We also require a projected separation of
\[
5 \,\mathrm{kpc} < \Delta r_{\mathrm{proj}} < 500 \,\mathrm{kpc}
\]
on the one hand to avoid confusion with internal galaxy substructures and on the other hand to limit the selection to realistic separations. Stellar mass ratios are defined as $\mu = M_2/M_1$, where $M_1$ is the mass of the primary galaxy and $M_2$ the secondary, and we adopt a threshold of $\mu \geq 0.1$.

To determine which galaxy pairs eventually merge, we use the merger tree on this initial basic selection.  
For pairs where the galaxies are in different snapshots, we trace the galaxy ID from the earlier snapshot (i.e., further back in time) forward through the merger tree until it reaches the snapshot of the companion galaxy. We then follow both galaxies together along the merger tree until their IDs coincide in the same snapshot, which indicates that the galaxies have merged.  For pairs where both galaxies are initially in the same snapshot, we perform the same procedure directly from that snapshot. If the galaxy IDs do not coincide by the last available snapshot, the pair is considered not to merge.

Building on this basic pair selection, we further apply an optimised set of dynamical criteria to refine the sample, using information on which pairs successfully merge. 
These cuts are redshift- and mass-dependent, and are defined in three dimensions using both spatial separation and relative velocity. To define the optimized selection limits, we examine the distribution of galaxy pairs in projected separation and velocity offset.
We decompose our galaxy sample into bins of primary galaxy stellar mass, mass ratio, and redshift.
For each bin, we define the upper limits of the projected separation ($\mathrm{\Delta}r_{\mathrm{proj,lim}}$) and projected velocity offset ($\mathrm{\Delta}v_{\mathrm{proj,lim}}$).

\begin{itemize}
    \item \textbf{Our criteria}
To identify optimal selection criteria for close galaxy pairs that are likely to merge, we searched for projected separation and velocity difference thresholds that maximize the classification performance within bins of stellar mass, mass ratio, and redshift. For each bin, we defined a rectangular selection region in the plane of projected distance ($\mathrm{\Delta r_{proj}}$) and velocity difference ($\mathrm{\Delta v_{proj}}$), and selected  pairs within regions of $\mathrm{\Delta r_{proj} }\leq \mathrm{\Delta}r_{\mathrm{proj,lim}}$ and $\mathrm{\Delta v_{proj} }\leq \mathrm{\Delta}v_{\mathrm{proj,lim}}$. The true merger classification for each pair was known from the analysis of the merger tree, which allowed us to evaluate the performance of each set of thresholds.

The classification results can be described using four quantities. True positives (TP) are pairs correctly classified as mergers. False positives (FP) are non-merging pairs that are incorrectly included in the selection box. False negatives (FN) are true mergers that are missed because they fall outside the selection limits. Finally, true negatives (TN) are non-merging pairs that are correctly excluded.

We defined purity and completeness as
\[
\mathrm{Purity} = \frac{TP}{TP + FP},
\qquad
\mathrm{Completeness} = \frac{TP}{TP + FN}.
\]
Purity measures the fraction of selected pairs that are true mergers, while completeness measures the fraction of all true mergers that are successfully selected. 

As a single figure of merit we adopted the Matthews Correlation Coefficient \citep[MCC][]{Pfister2020}, which incorporates all four terms of the confusion matrix:
\[
\mathrm{MCC} = \rm \frac{TP \cdot TN - FP \cdot FN}
{\sqrt{(TP+FP)(TP+FN)(TN+FP)(TN+FN)}}.
\]
The MCC ranges from $-1$ (complete misclassification) to $+1$ (perfect classification), with 0 corresponding to random selection. Unlike purity or completeness alone, the MCC provides a balanced evaluation of performance and remains robust under class imbalance, which is critical because true mergers constitute only a small fraction of all pairs.

In our analysis, we used a precision of $10^{-4}$ when computing the MCC in each mass, mass ratio and redshift bins. This means that, after calculating the MCC values for all bins, we search for the highest MCC and then consider bins within $10^{-4}$ of this maximum. Among these, we select the 
[$\mathrm{\Delta}r_{\rm proj}$, $\mathrm{\Delta}v_{\rm proj}$] bin containing the largest number of galaxies to define our final criteria.

For each mass, mass ratio, and redshift bin we performed a search over a grid of $\mathrm{\Delta}r$ and $\mathrm{\Delta}v$ limits, evaluated the MCC for every candidate box, and selected the combination of $(\mathrm{\Delta}r_{\mathrm{lim}}, \mathrm{\Delta}v_{\mathrm{lim}})$ that maximized the coefficient. To avoid pathological solutions, we required each bin to contain a minimum number of sources and each candidate box to enclose a minimum number of pairs. In cases where multiple boxes yielded nearly identical MCC values, we chose the box that contained the larger number of pairs. We calculate limits only for bins where we have more than 10 pairs (Table \ref{tab:dvdr}). 

\begin{table*}
\centering
\begin{tabular}{ccccccccc}
\hline \hline\hline
$\log_{10}\frac{M_{\mathrm{gal,1}}}{M_\odot}$ & Mass ratio & Redshift & $\mathrm{\Delta r_{proj,lim}}$ [pKpc] & $\mathrm{\Delta v_{proj,lim}}$ [km/s] & MCC & Purity & Completeness & $\mathrm{N_{pairs,sel}}$ \\
\hline \hline\hline
\mbox{[9,9.5]} & $\mu>0.25$ & \mbox{[0,1]} & 105 & 449 & 0.523 & 0.374 & 0.852 & 1581 \\
\mbox{[9,9.5]} & $\mu>0.25$ & \mbox{[1,2]} & 60 & 152 & 0.580 & 0.468 & 0.809 & 5002\\
\mbox{[9,9.5]} & $\mu>0.25$ & \mbox{[2,3]} & 141 & 86 & 0.797 & 0.753 & 0.923 & 8789\\
\mbox{[9,9.5]} & $\mu>0.25$ & \mbox{[3,4]} & 240 & 77 & 0.873 & 0.871 & 0.941 & 6357\\
\mbox{[9,9.5]} & $\mu>0.25$ & \mbox{[4,5]} & 252 & 122 & 0.867 & 0.907 & 0.912 & 2116\\
\mbox{[9,9.5]} & $\mu>0.25$ & \mbox{[5,6]} & 369 & 161 & 0.842 & 0.896 & 0.909 & 211\\
\hline\hline
\mbox{[9.5,10]} & $\mu<0.25$ & \mbox{[0,1]} & 102 & 386 & 0.593 & 0.465 & 0.893 & 1031\\
\mbox{[9.5,10]} & $\mu<0.25$ & \mbox{[1,2]} & 69 & 107 & 0.635 & 0.541 & 0.851 & 2165\\
\mbox{[9.5,10]} & $\mu<0.25$ & \mbox{[2,3]} & 150 & 80 & 0.851 & 0.841 & 0.941 & 3632\\
\mbox{[9.5,10]} & $\mu<0.25$ & \mbox{[3,4]} & 261 & 110 & 0.901 & 0.908 & 0.968 & 1851\\
\mbox{[9.5,10]} & $\mu<0.25$ & \mbox{[4,5]} & 366 & 188 & 0.904 & 0.928 & 0.976 & 430\\
\mbox{[9.5,10]} & $\mu<0.25$ & \mbox{[5,6]} & 366 & 398 & 0.946 & 0.956 & 1.000 & 45\\
\hline
\mbox{[9.5,10]} & $\mu>0.25$ & \mbox{[0,1]} & 111 & 401 & 0.589 & 0.453 & 0.900 & 3098\\
\mbox{[9.5,10]} & $\mu>0.25$ & \mbox{[1,2]} & 63 & 128 & 0.646 & 0.568 & 0.841 & 7712\\
\mbox{[9.5,10]} & $\mu>0.25$ & \mbox{[2,3]} & 153 & 86 & 0.843 & 0.829 & 0.939 & 11067\\
\mbox{[9.5,10]} & $\mu>0.25$ & \mbox{[3,4]} & 222 & 95 & 0.908 & 0.920 & 0.962 & 6346\\
\mbox{[9.5,10]} & $\mu>0.25$ & \mbox{[4,5]} & 309 & 152 & 0.884 & 0.929 & 0.950 & 1527\\
\mbox{[9.5,10]} & $\mu>0.25$ & \mbox{[5,6]} & 309 & 164 & 0.864 & 0.935 & 0.923 & 153\\
\hline\hline
\mbox{[10,10.5]} & $\mu<0.25$ & \mbox{[0,1]} & 114 & 386 & 0.648 & 0.544 & 0.918 & 1876\\
\mbox{[10,10.5]} & $\mu<0.25$ & \mbox{[1,2]} & 63 & 188 & 0.684 & 0.630 & 0.877 & 4375\\
\mbox{[10,10.5]} & $\mu<0.25$ & \mbox{[2,3]} & 174 & 122 & 0.877 & 0.873 & 0.971 & 4899\\
\mbox{[10,10.5]} & $\mu<0.25$ & \mbox{[3,4]} & 294 & 131 & 0.927 & 0.945 & 0.978 & 1537\\
\mbox{[10,10.5]} & $\mu<0.25$ & \mbox{[4,5]} & 342 & 167 & 0.901 & 0.965 & 0.956 & 202\\
\mbox{[10,10.5]} & $\mu<0.25$ & \mbox{[5,6]} & 294 & 215 & 0.933 & 1.000 & 0.958 & 23\\
\hline
\mbox{[10,10.5]} & $\mu>0.25$ & \mbox{[0,1]} & 111 & 407 & 0.648 & 0.554 & 0.908 & 1798\\
\mbox{[10,10.5]} & $\mu>0.25$ & \mbox{[1,2]} & 72 & 152 & 0.707 & 0.659 & 0.893 & 4478\\
\mbox{[10,10.5]} & $\mu>0.25$ & \mbox{[2,3]} & 210 & 119 & 0.879 & 0.883 & 0.966 & 3899\\
\mbox{[10,10.5]} & $\mu>0.25$ & \mbox{[3,4]} & 300 & 119 & 0.923 & 0.953 & 0.977 & 1138\\
\mbox{[10,10.5]} & $\mu>0.25$ & \mbox{[4,5]} & 405 & 182 & 0.973 & 1.000 & 0.983 & 117\\
\mbox{[10,10.5]} & $\mu>0.25$ & \mbox{[5,6]} & 402 & 209 & 1.000 & 1.000 & 1.000 & 11\\
\hline\hline
\mbox{[10.5,11]} & $\mu<0.25$ & \mbox{[0,1]} & 147 & 434 & 0.702 & 0.643 & 0.951 & 967\\
\mbox{[10.5,11]} & $\mu<0.25$ & \mbox{[1,2]} & 99 & 224 & 0.724 & 0.707 & 0.935 & 2032\\
\mbox{[10.5,11]} & $\mu<0.25$ & \mbox{[2,3]} & 198 & 191 & 0.902 & 0.924 & 0.981 & 1046\\
\mbox{[10.5,11]} & $\mu<0.25$ & \mbox{[3,4]} & 480 & 158 & 0.934 & 0.959 & 1.000 & 146\\
\mbox{[10.5,11]} & $\mu<0.25$ & \mbox{[4,5]} & 417 & 95 & 1.000 & 1.000 & 1.000 & 10\\
\hline
\mbox{[10.5,11]} & $\mu>0.25$ & \mbox{[0,1]} & 192 & 482 & 0.691 & 0.619 & 0.939 & 1031\\
\mbox{[10.5,11]} & $\mu>0.25$ & \mbox{[1,2]} & 96 & 347 & 0.710 & 0.693 & 0.901 & 1779\\
\mbox{[10.5,11]} & $\mu>0.25$ & \mbox{[2,3]} & 342 & 203 & 0.853 & 0.864 & 0.987 & 939\\
\mbox{[10.5,11]} & $\mu>0.25$ & \mbox{[3,4]} & 489 & 128 & 0.972 & 0.977 & 1.000 & 88\\
\hline\hline
\mbox{[11,11.5]} & $\mu<0.25$ & \mbox{[0,1]} & 240 & 422 & 0.723 & 0.732 & 0.960 & 331\\
\mbox{[11,11.5]} & $\mu<0.25$ & \mbox{[1,2]} & 126 & 347 & 0.717 & 0.757 & 0.954 & 407\\
\mbox{[11,11.5]} & $\mu<0.25$ & \mbox{[2,3]} & 252 & 209 & 0.852 & 0.920 & 0.979 & 100\\
\hline
\mbox{[11,11.5]} & $\mu>0.25$ & \mbox{[0,1]} & 363 & 425 & 0.683 & 0.656 & 0.962 & 258\\
\mbox{[11,11.5]} & $\mu>0.25$ & \mbox{[1,2]} & 165 & 344 & 0.720 & 0.765 & 0.921 & 226\\
\mbox{[11,11.5]} & $\mu>0.25$ & \mbox{[2,3]} & 312 & 362 & 0.856 & 0.921 & 0.986 & 76\\
\hline\hline\hline
\end{tabular}
\caption{Selection criteria and statistics for galaxy pairs in our sample.}
\label{tab:dvdr}
\end{table*}

\begin{figure}
    \centering
    \includegraphics[width=0.48\textwidth]{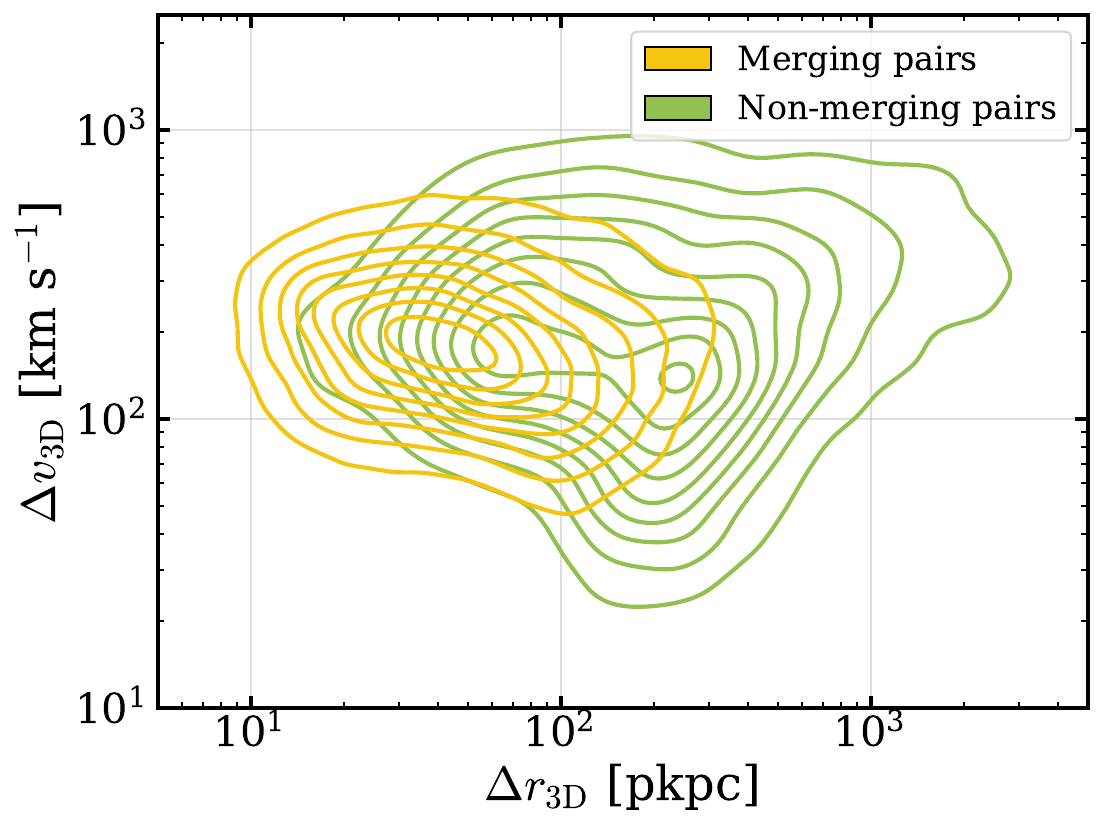}
    \caption{Distribution of galaxy pairs in 3D separation and velocity space ($\Delta r_{\rm 3D}$, $\Delta v_{\rm 3D}$) with log-scaled axes. Contours show merging pairs in yellow and non-merging pairs in green, indicating regions of higher pair density. Merging pairs cluster at small 3D separations and low velocity differences, as expected for gravitationally bound systems, though mergers can still occur at separations up to $\sim1000$ kpc. The broad overlap between merging and non-merging pair distributions underscores why simple distance cuts alone are insufficient to cleanly separate the two populations.
}
    \label{dv3d}
\end{figure}

    \item \textbf{\citet{Puskas2025}}

We compared this approach to other approaches used in the literature. \citet{Puskas2025} to determine whether two galaxies are consistent in redshift, use a probabilistic approach based on the posterior distribution of the photometric redshifts for two galaxies, but also include spectroscopic redshifts. Given that we have the exact redshift for each galaxy in the lightcone of Horizon-AGN, we adopt the latter.  In their framework, each spectroscopic redshift measurement is modelled as a Gaussian probability distribution \(P(z\,|\,z_i,\sigma_z)\) centred on the reported redshift \(z_i\), with a fixed width \(\sigma_z = 0.01\).  For each pair with redshifts \(z_1\) and \(z_2\), we compute the overlap function
\[
Z(z) = \frac{2\,P_1(z)\,P_2(z)}{P_1(z)+P_2(z)},
\]

on a sufficiently wide redshift grid and evaluate the integrated overlap
\[
N_z = \int Z(z)\,\mathrm{d}z.
\]

We classify pairs as redshift-consistent when \(N_z > 0.8\).

The probabilistic redshift requirement is applied together with the standard projected-separation and mass ratio cuts used in \cite{Puskas2025}:
\[
5 \mathrm{kpc} \le \Delta r_{\mathrm{proj}} \le 30\,\mathrm{kpc}, \qquad
\mu \ge 0.25, \qquad N_z > 0.8.
\]

For Figure~\ref{dvdrlim}, we define $v_{\mathrm{proj}}$ limit as the maximum projected velocity difference among all galaxy pairs within each redshift bin.

    \item \textbf{\citet{Pfister2020}} 
   This work also used the Horizon-AGN simulation, but they limited their analysis to $z<1$. We here consider all available redshifts, and thus extend the analysis to high redshift.   In  \citet{Pfister2020}  a stellar mass-dependent redshift proximity criterion is proposed:
    \[
    \Delta z < 6 \times 10^{-4} \left(\frac{M_1}{10^{10}\,\msun}\right)^{0.28},
    \]
    along with a projected separation cut scaled by stellar mass:
    \[
    \Delta r_{\mathrm{proj}} < 84 \left(\frac{M_1}{10^{10}\,\msun}\right)^{0.2}\,\mathrm{kpc}.
    \]

    \item \textbf{\citet{Ventou2019}} 
    
In this paper a two-tiered selection based on projected separation and line-of-sight velocity difference is used:
    \[
    \begin{cases}
    5 \leq \Delta r_{\mathrm{proj}} \leq 50\,\mathrm{kpc}, & \Delta v_{\mathrm{proj}} \leq 300\,\mathrm{km\,s^{-1}}, \\
    50 < \Delta r_{\mathrm{proj}} \leq 100\,\mathrm{kpc}, & \Delta v_{\mathrm{proj}} \leq 100\,\mathrm{km\,s^{-1}}.
    \end{cases}
    \]
\end{itemize}

The line-of-sight velocity difference between galaxies is computed as:
\[
\Delta v_{\mathrm{proj}} = c \frac{|z_1 - z_2|}{1 + (z_1 + z_2)/2},
\]
where $c$ is the speed of light. This formula allows consistent comparison to observational limits such as the commonly adopted threshold of $\Delta v_{\mathrm{proj}} \leq 500\,\mathrm{km\,s^{-1}}$ \citep[e.g.][]{Patton2000,DePropris2007}.

Figure~\ref{dvdrlim} shows the limits in projected separation ($\Delta r_{\rm proj}$) and projected velocity difference ($\Delta v_{\rm proj}$) used to define close galaxy pairs in our sample, for different stellar mass, mass ratio, and redshift bins. As expected, more massive galaxies tend to have larger $\Delta r_{\rm proj}$ limits, reflecting the fact that physically larger systems can interact over wider projected separations. Compared to previous studies \citep[e.g.,][]{Puskas2025,Pfister2020,Ventou2019}, our limits generally cover a broader range in projected separation, allowing the selection of galaxy pairs at higher distances. While \citet{Pfister2020} also employed MCC, their analysis was restricted to $z<1$, whereas our selection extends to higher redshifts.

Figure~\ref{dv3d} presents the distribution of galaxy pairs in three-dimensional separation and velocity space ($\Delta r_{\rm 3D}$, $\Delta v_{\rm 3D}$). Pairs that eventually merge are shown as yellow contours and non-merging pairs as green. As expected, galaxy pairs that do not merge tend to occupy larger 3D separations, while merging pairs are concentrated at smaller distances. Nevertheless, some mergers occur even up to $\sim 1000$\,kpc, highlighting that long-range interactions can still result in mergers, and emphasizing the importance of defining dynamical limits that capture the full range of physically relevant separations and velocities.

Together, these figures show that our adaptive, bin-dependent limits in projected separation and velocity allow for a more flexible and physically motivated selection of galaxy pairs compared to previous fixed criteria, while remaining consistent with observed trends in galaxy mass and redshift.

\section{Results}
\label{sec:results}

In the following sections we present our results, while in the Appendix we collect figures that can be used to apply our criteria to observational samples.

\subsection{Galaxy Pair Fraction}

\begin{figure*}[tb]
    \centering

    \begin{minipage}{\textwidth}
        \centering
        \vspace{-6cm}
        \includegraphics[width=\textwidth]{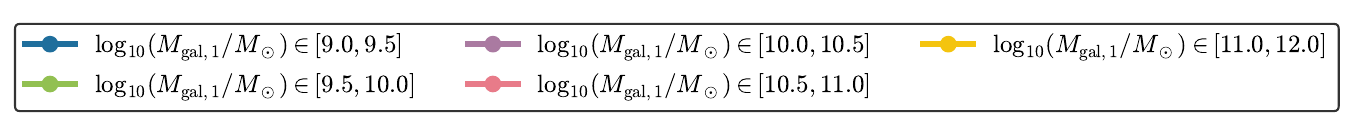}
    \end{minipage}

    \vspace{-2cm} 

    \begin{minipage}{\textwidth}
        \centering
        \begin{subfigure}[t]{0.47\textwidth}
            \centering
            \includegraphics[width=\textwidth]{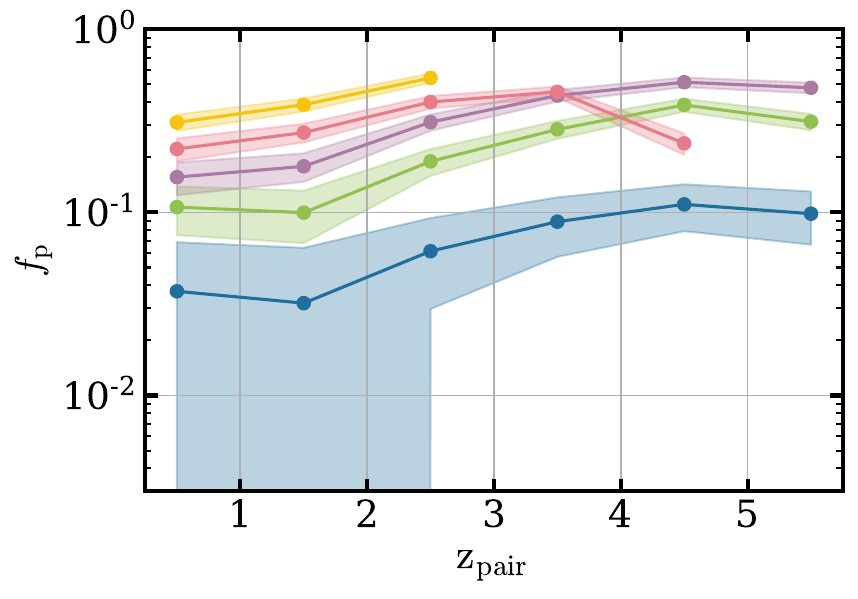}
            \caption{Our criteria.}
            \label{fp_kate}
        \end{subfigure}
        \hfill
        \begin{subfigure}[t]{0.47\textwidth}
            \centering
            \includegraphics[width=\textwidth]{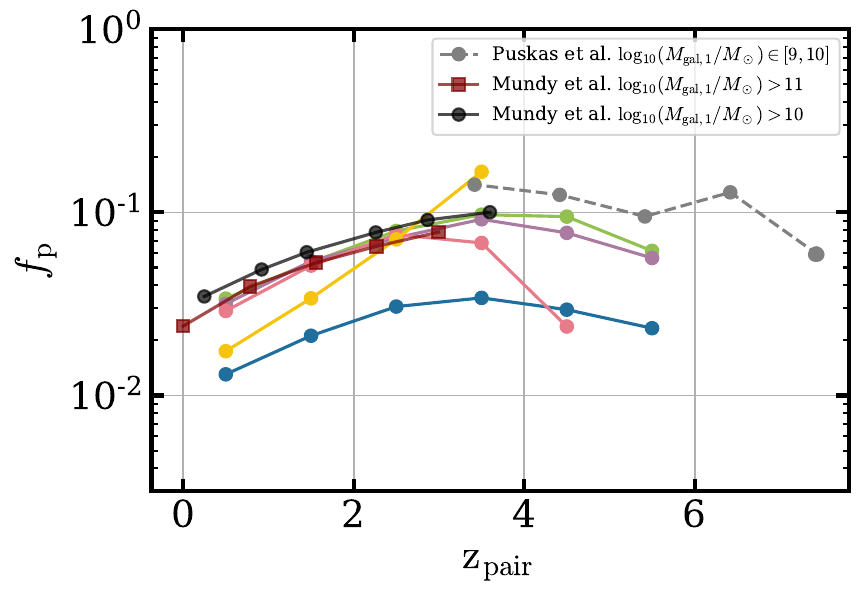}
            \caption{Selection criteria from \cite{Puskas2025, Mundy2017}}
            \label{fp_puskas}
        \end{subfigure}

        \vspace{0.3cm}

        \begin{subfigure}[t]{0.47\textwidth}
            \centering
            \includegraphics[width=\textwidth]{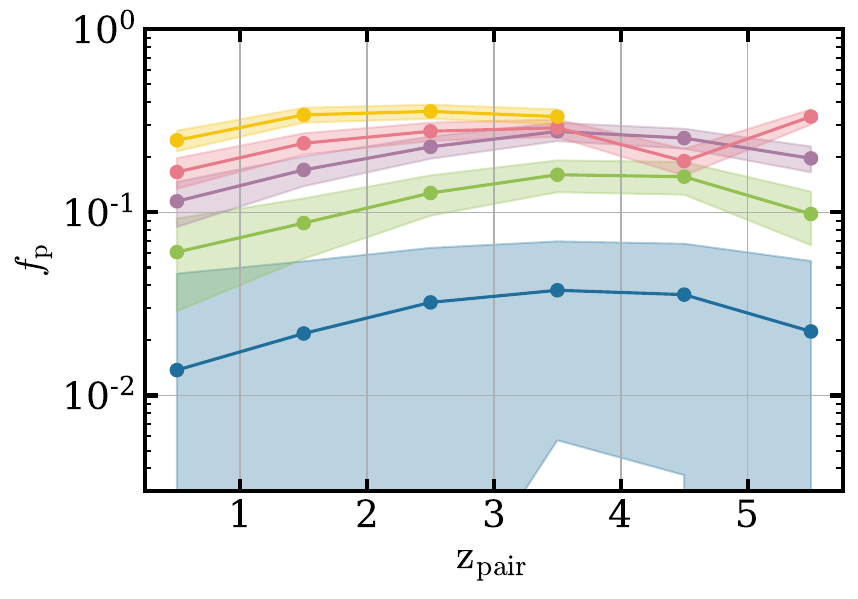}
            \caption{Selection criteria from \cite{Pfister2020}}
            \label{fp_pfister}
        \end{subfigure}
        \hfill
        \begin{subfigure}[t]{0.47\textwidth}
            \centering
            \includegraphics[width=\textwidth]{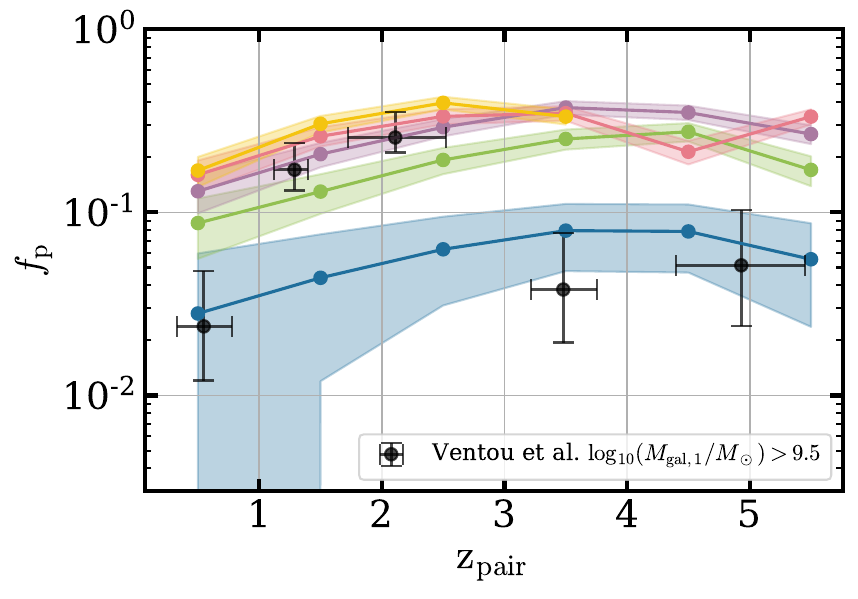}
            \caption{Selection criteria from \cite{Ventou2019}}
            \label{fp_ventou}
        \end{subfigure}
    \end{minipage}

    \caption{Galaxy pair fraction as a function of redshift and stellar mass. Each color represents a different mass bin, and each panel shows the pair fraction from our sample (\ref{fp_kate}) compared to the criteria from the literature (\ref{fp_puskas}, \ref{fp_pfister}, \ref{fp_ventou}). Uncertainties are shown as the corresponding filled color, calculated from Poisson statistics (except for panel~\ref{fp_puskas}, where errorbars are omitted for clarity). Regardless of the selection method used, more massive galaxies consistently show higher pair fractions, driven by their denser environments and greater likelihood of having nearby companions.}
    \label{fp}
\end{figure*}


We apply the approaches presented in Section~\ref{sec:pairselection} to the full simulated pair sample and compare pair fractions using the bin-optimized approach to those obtained when applying the selection criteria proposed in \citet{Puskas2025}, \citet{Pfister2020}, \citet{Ventou2019}, and \citet{Conselice2022}.

We calculate the galaxy pair fraction in bins of stellar mass and redshift, defined as
\begin{equation}
f_{\mathrm{p}} = \frac{N_{\mathrm{pair}}}{N_{\mathrm{total}}},
\label{eq:fp}
\end{equation}
where $N_{\mathrm{pair}}$ is the number of galaxies that are part of selected pairs and $N_{\mathrm{total}}$ is the total number of galaxies in the lightcone within the corresponding redshift--mass bin.

The resulting pair fractions are shown in Fig.~\ref{fp} and Fig.~\ref{histfp} in the Appendix. Figure~\ref{fp_kate} displays the pair fraction obtained with our selection criteria described in Section~\ref{sec:methods}, while Figs.~\ref{fp_puskas}, \ref{fp_pfister}, \ref{fp_ventou} show the corresponding results when applying the methods of \citet{Puskas2025}, \citet{Pfister2020}, and \citet{Ventou2019}, respectively. In panel~\ref{fp_puskas} the results are also compared to the observational analyses from \citet{Puskas2025} and \citet{Mundy2017}. 

 As shown in Fig.~\ref{fp}, panels~\ref{fp_kate},~\ref{fp_pfister}, and~\ref{fp_ventou}, all exhibit the theoretically expected increase of pair fraction with stellar mass. This trend reflects both the increasing interactions of massive galaxies, which are hosted in dense environment, and their higher probability of hosting multiple companions. The drop at $\log (M_{1}/\msun)\in[10.5,11]$ is driven by low-number statistics at $z\in[4,5]$, where the lightcone contains fewer galaxies overall, while the lowest mass bin ($10^9\mbox{--}10^{9.5} \, \msun$) starts being incomplete as galaxies get close to the resolution limit. 

Panel~\ref{fp_puskas}, which represents observationally motivated cuts from \citet{Puskas2025} and \citet{Conselice2022}, compares to observational analyses in different mass and redshift regimes from the same papers.  The pair fractions we obtain for $\log(M_{1}/\msun)\in[9.5,10]$ (our lowest mass bin suffers from incompleteness being close to the simulation's resolution) lie within the range reported in \citet{Puskas2025}. We note that using the same selection as \citet{Puskas2025} we find a weaker trend with galaxy mass, contrary to the other selections shown in panels~\ref{fp_kate},~\ref{fp_pfister}, and~\ref{fp_ventou}. This is in broad agreement with \citet{Puskas2025}, where little difference is found for different mass bins with $\log(M_{1}/\msun)\in[8.5,10]$, but we do not reproduce their trends of increasing pair fraction with decreasing mass.  When comparing our results to \citet{Ventou2019}, using their criteria, we find comparable results up to $z=2$, but a higher pair fraction at higher redshift, similar to what is reported in \citet{Ventou2019} for the EAGLE simulation. 

When comparing different selection criteria, we see that those that are simulation-based and optimized on selecting pairs that have the highest chance of merging, yield systematically higher pair fractions in the same mass bins (\ref{fp_kate},~\ref{fp_pfister}, and~\ref{fp_ventou}). This mass and redshift-dependent limits, include a larger number of physically associated companions, which although at large physical separation are indeed going to merge. Unexpectedly, separations as large as $300\mbox{--}400$\,kpc can lead to a high fraction of galaxies effectively merging, as long as the time allowed for the merger is large (in other words, if $z_{\rm pair}$ is sufficiently high). Overall, these comparisons demonstrate that fixed or weakly evolving selection criteria can underestimate pair fractions.


\subsection{Galaxy Pair Merger Fractions and Timescales}

To determine which galaxy pairs eventually merge, we use the merger tree. For pairs where the galaxies are in different snapshots, we trace the galaxy ID from the earlier snapshot (i.e., further back in time) forward through the merger tree until it reaches the snapshot of the companion galaxy. We then follow both galaxies together along the merger tree until their IDs coincide in the same snapshot, which indicates that the galaxies have merged.  For pairs where both galaxies are initially in the same snapshot, we perform the same procedure directly from that snapshot. If the galaxy IDs do not coincide by the last available snapshot, the pair is considered not to merge.

\begin{figure*}[tb]
    \centering

    \begin{minipage}{\textwidth}
        \centering
        \includegraphics[width=\textwidth]{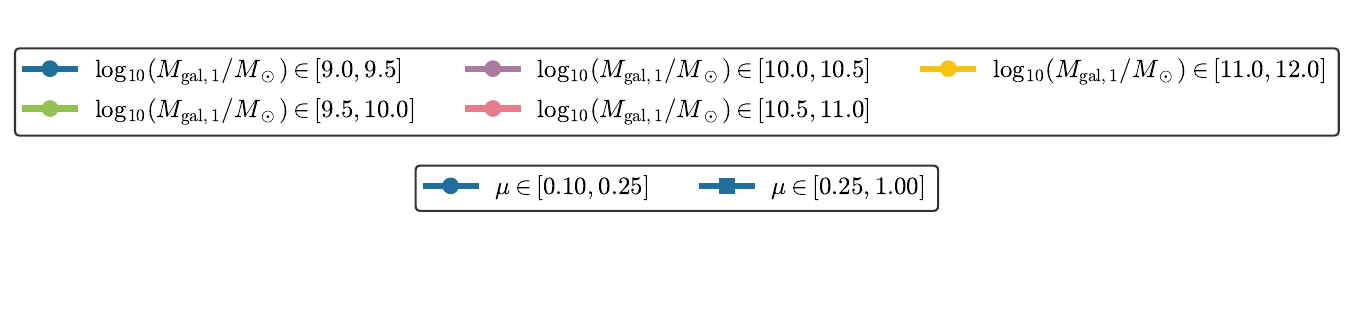}
    \end{minipage}

    \vspace{-1cm} 

    \begin{minipage}{\textwidth}
        \centering
        \begin{subfigure}[t]{0.49\textwidth}
            \centering
            \includegraphics[width=\textwidth]{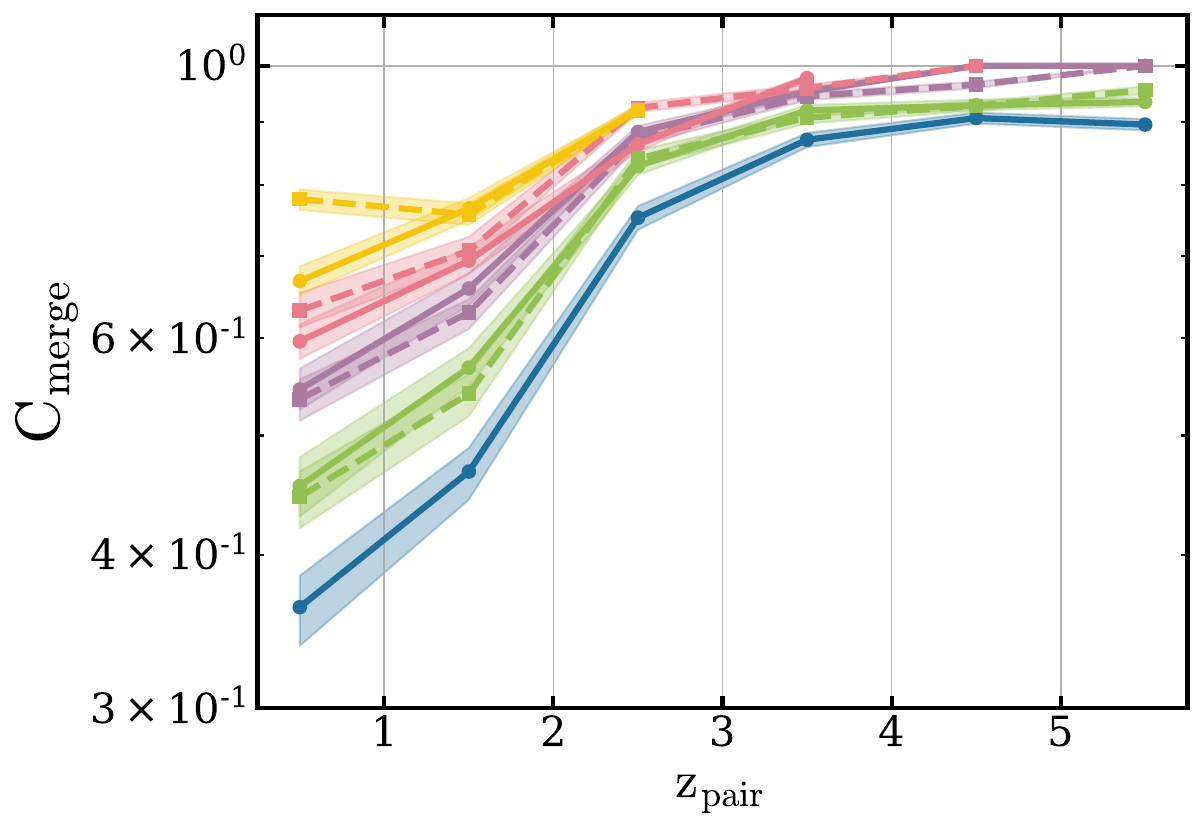}
            \caption{Our criteria.}
            \label{cmerge_kate}
        \end{subfigure}
        \hfill
        \begin{subfigure}[t]{0.49\textwidth}
            \centering
            \includegraphics[width=\textwidth]{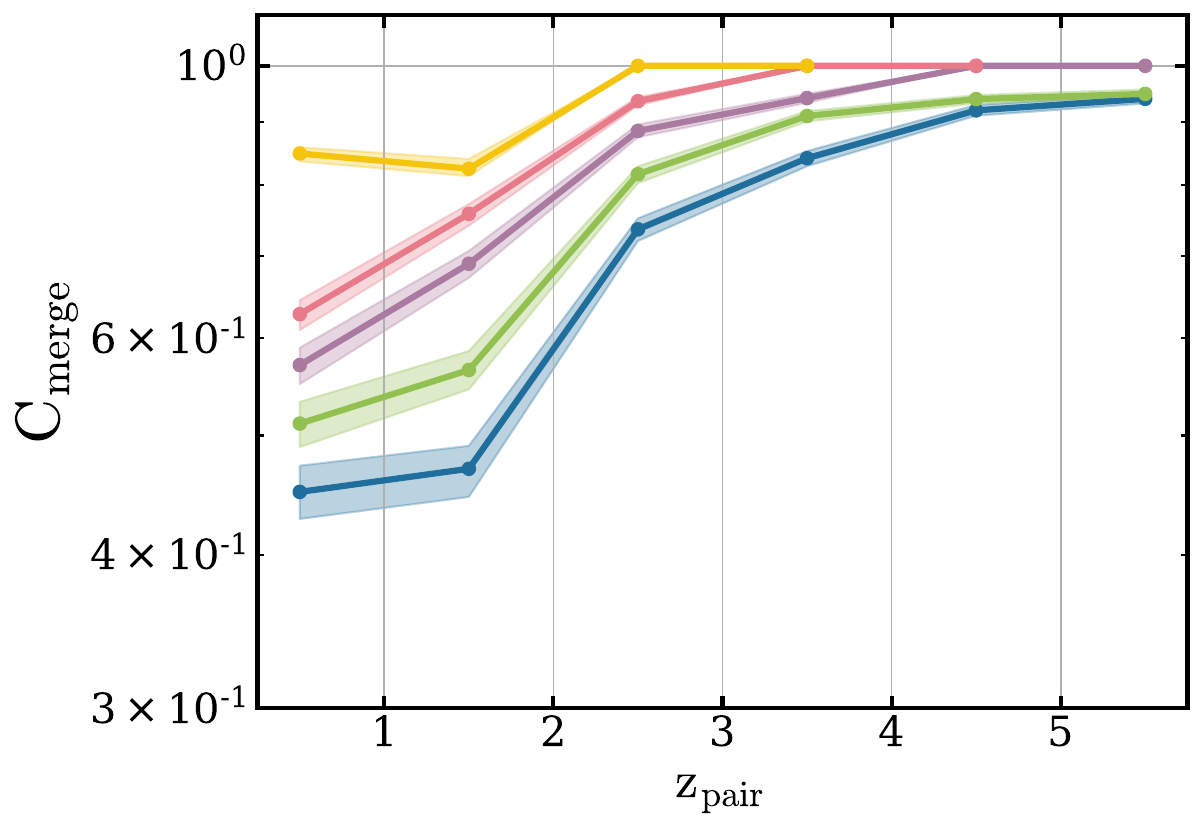}
            \caption{Selection criteria from \cite{Puskas2025}}
            \label{cmerge_puskas}
        \end{subfigure}

        \vspace{0.3cm}

        \begin{subfigure}[t]{0.49\textwidth}
            \centering
            \includegraphics[width=\textwidth]{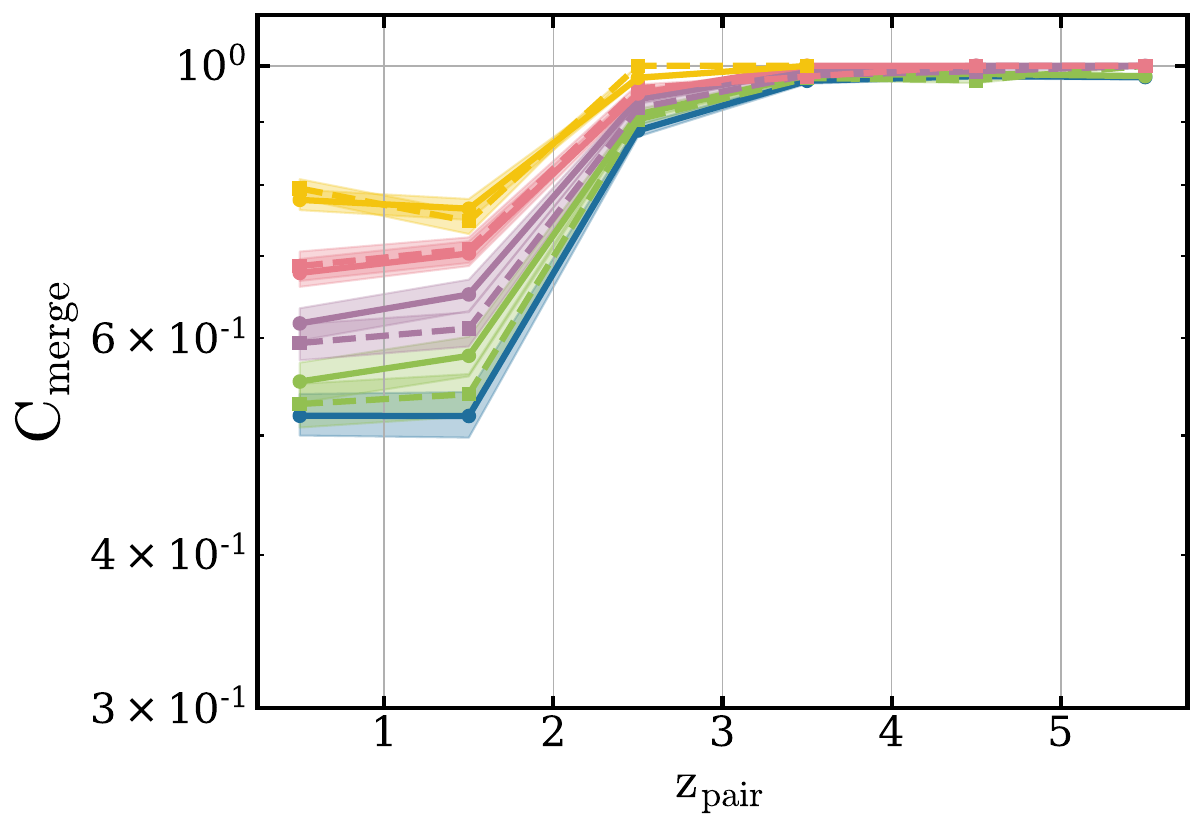}
            \caption{Selection criteria from \cite{Pfister2020}}
            \label{cmerge_pfister}
        \end{subfigure}
        \hfill
        \begin{subfigure}[t]{0.49\textwidth}
            \centering
            \includegraphics[width=\textwidth]{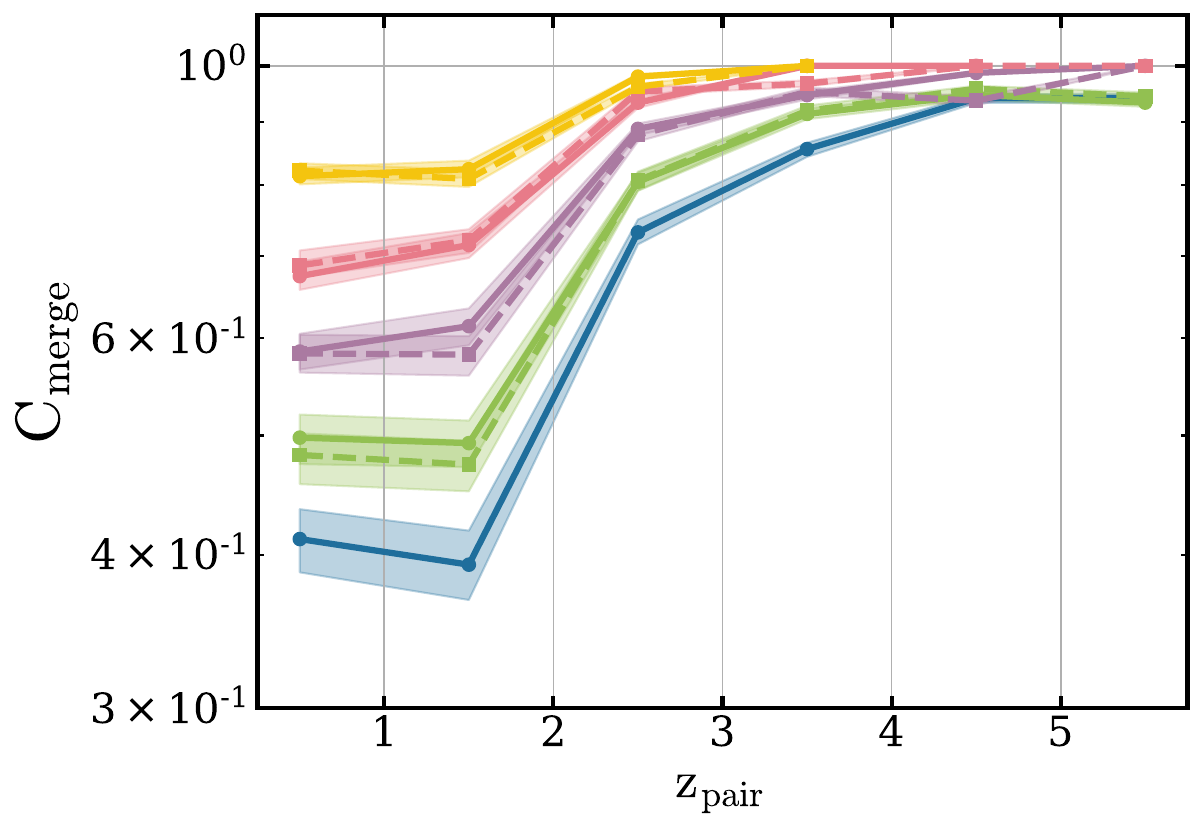}
            \caption{Selection criteria from \cite{Ventou2019}}
            \label{cmerge_ventou}
        \end{subfigure}
    \end{minipage}

    \caption{Galaxy merger fraction -- the fraction of selected pairs that successfully merge -- as a function of redshift and stellar mass. Each color represents a different mass bin, with solid lines showing major mergers and dashed lines showing minor mergers. Uncertainties were calculated using Poisson statistics. Panel \ref{cmerge_kate} shows results using the criteria from this paper, while \ref{cmerge_puskas}, \ref{cmerge_pfister}, and \ref{cmerge_ventou} show results using criteria from the literature. All selection criteria show increase of galaxy merger fraction with primary galaxy mass and redshift. Pairs of massive galaxies are more likely to complete a merger, and pairs at higher redshift have had more time to merge.}
    \label{cmerge}
\end{figure*}

\begin{figure*}[tb]
    \centering

    \begin{minipage}{\textwidth}
        \centering
        \includegraphics[width=\textwidth]{images/legend_mass_and_mu.pdf}
    \end{minipage}

    \vspace{-1cm}

    \begin{minipage}{\textwidth}
        \centering
        \begin{subfigure}[t]{0.49\textwidth}
            \centering
            \includegraphics[width=\textwidth]{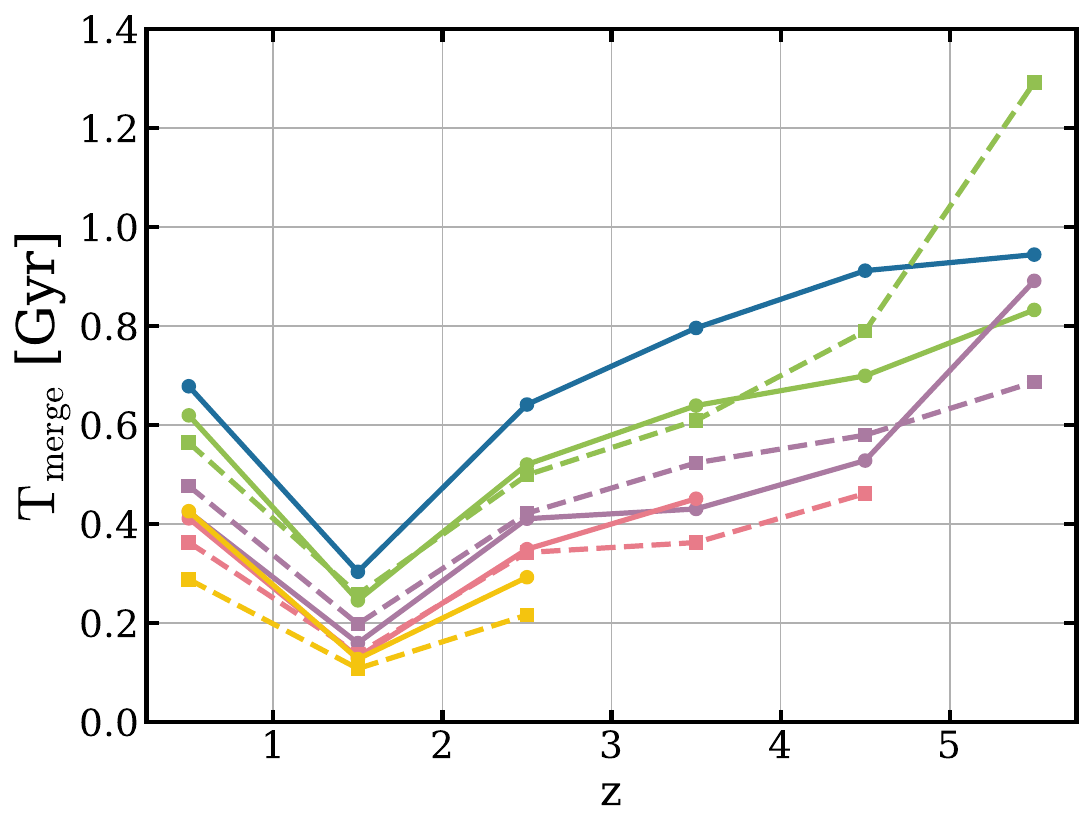}
            \caption{Our criteria.}
            \label{tmerge_kate}
        \end{subfigure}
        \hfill
        \begin{subfigure}[t]{0.49\textwidth}
            \centering
            \includegraphics[width=\textwidth]{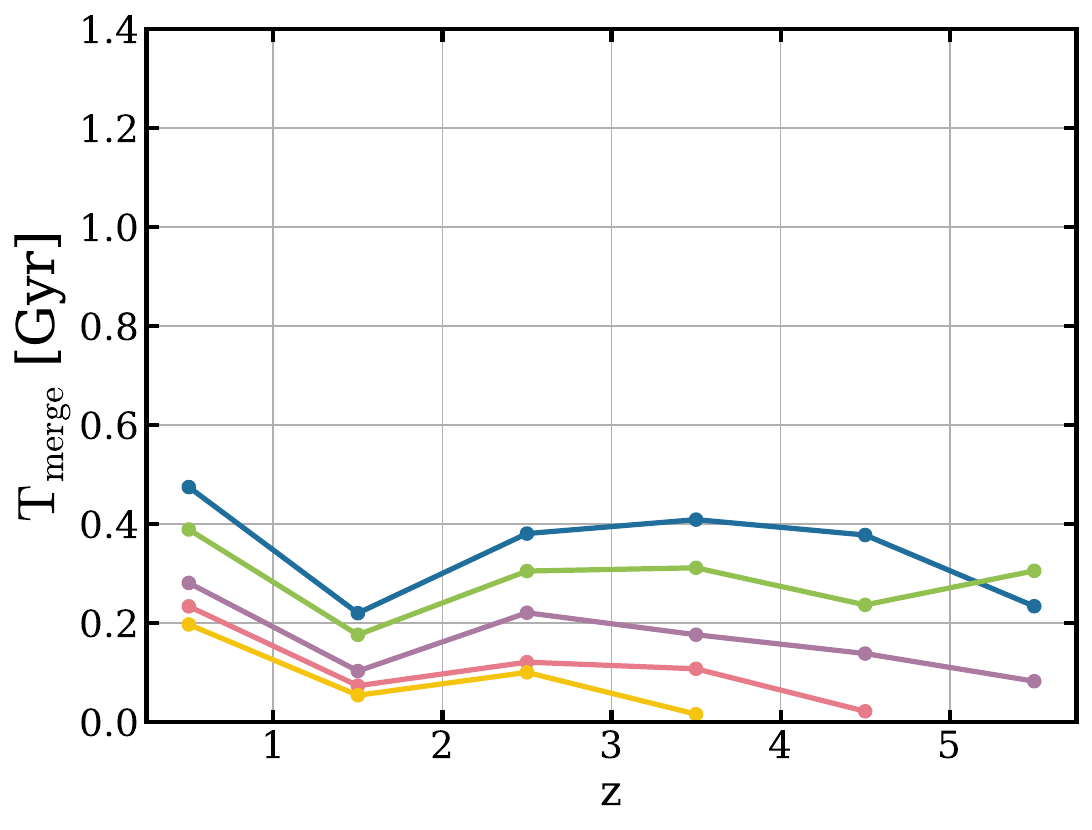}
            \caption{Selection criteria from \cite{Puskas2025}}
            \label{tmerge_puskas}
        \end{subfigure}

        \vspace{0cm}

        \begin{subfigure}[t]{0.49\textwidth}
            \centering
            \includegraphics[width=\textwidth]{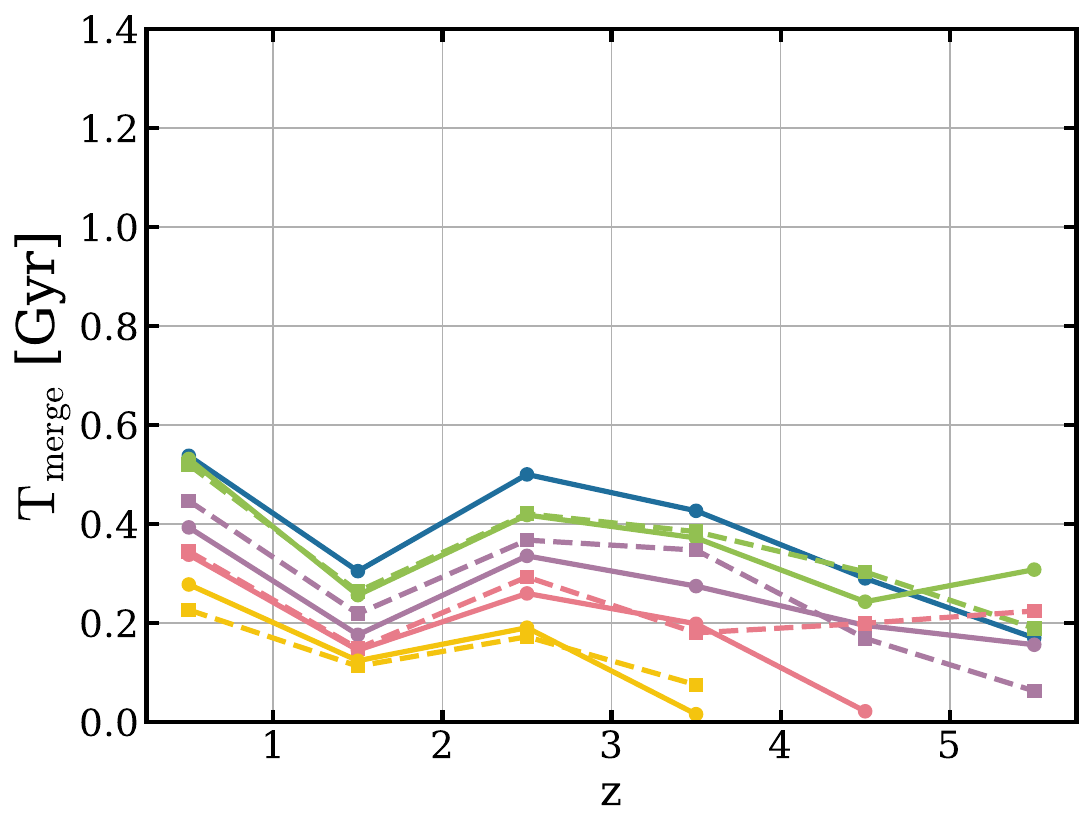}
            \caption{Selection criteria from \cite{Pfister2020}}
            \label{tmerge_pfister}
        \end{subfigure}
        \hfill
        \begin{subfigure}[t]{0.49\textwidth}
            \centering
            \includegraphics[width=\textwidth]{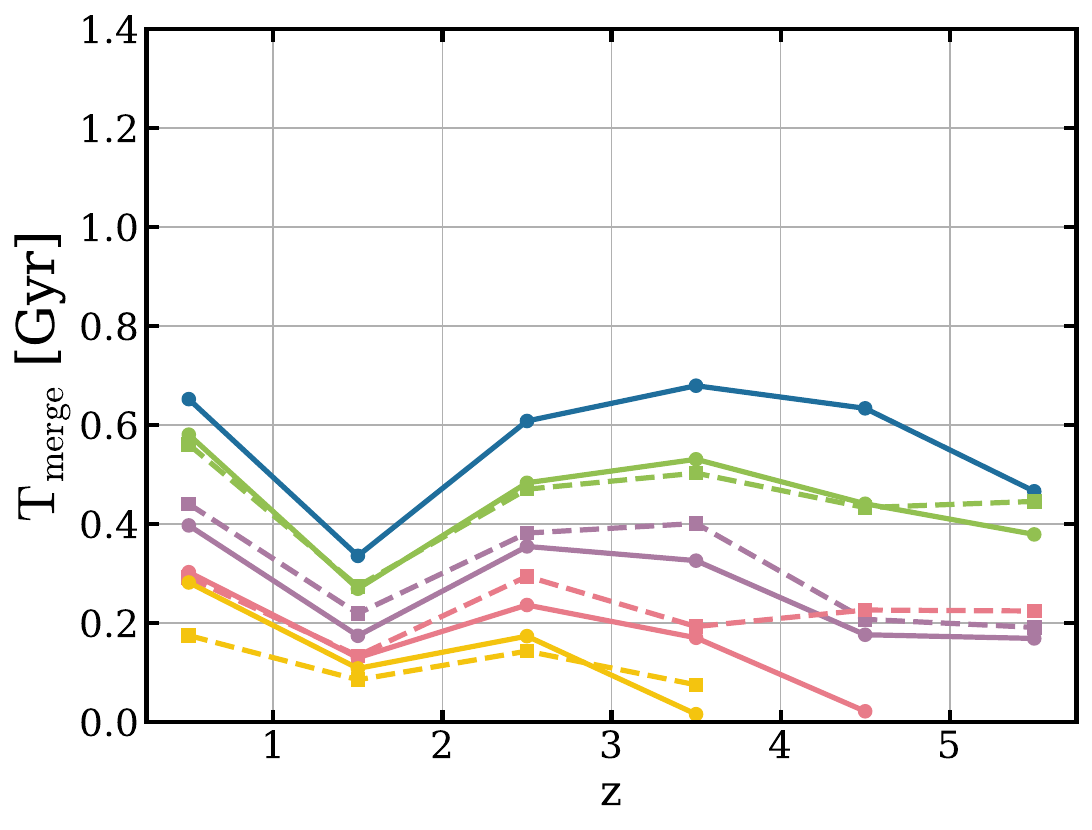}
            \caption{Selection criteria from \cite{Ventou2019}}
            \label{tmerge_ventou}
        \end{subfigure}
    \end{minipage}

    \caption{Galaxy average merger timescale as a function of redshift and stellar mass. Each color represents a different mass bin, with solid lines for major mergers and dashed lines for minor mergers. Uncertainties were calculated using Poisson statistics. Panel \ref{tmerge_kate} shows results using the criteria from this paper, while \ref{tmerge_puskas}, \ref{tmerge_pfister}, and \ref{tmerge_ventou} show results using criteria from the literature. In all selection criteria, we observe that the galaxy merger timescale increases with decreasing primary galaxy mass, as smaller galaxies take longer to merge than more massive ones. Our method includes more widely separated pairs, which leads to longer average merger times. The noticeable drop in the timescale at $z=1.5$ is caused by the introduction of a new refinement level in the simulation at $z=1.5$.}
    \label{tmerge}
\end{figure*}

\begin{figure*}[tb]
    \centering
    \includegraphics[width=\textwidth]{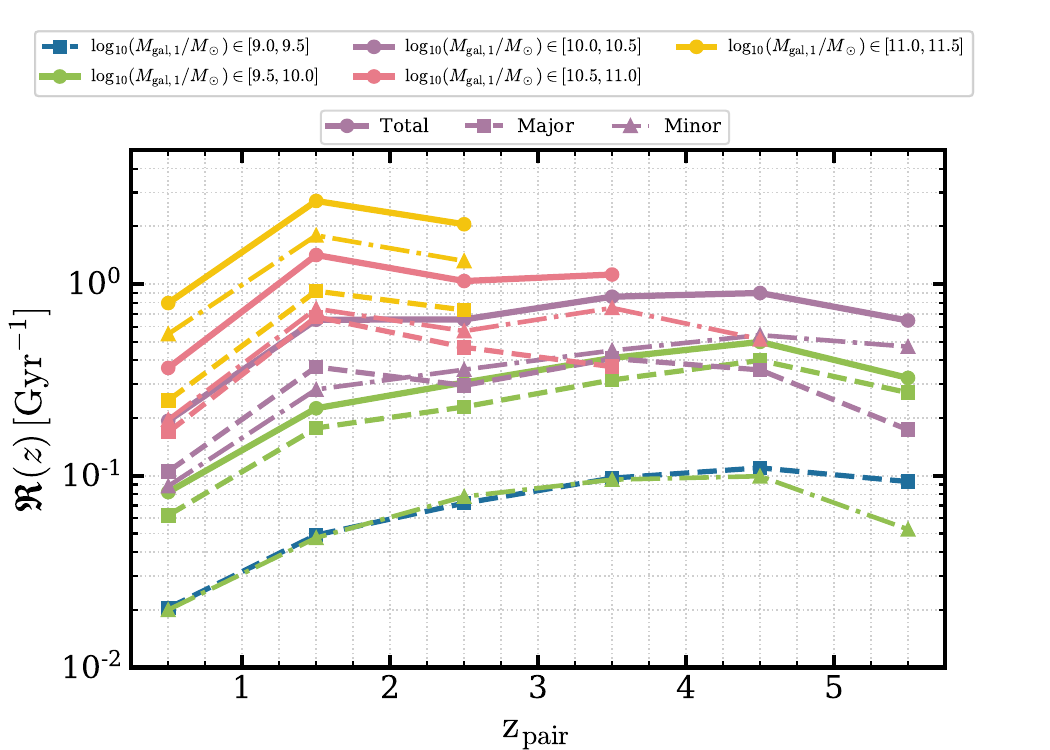}
    \caption{Galaxy merger rate calculated from our sample. Each color represents a different mass bin. The solid line with circle markers sho
ws the total merger rate (major + minor), the dashed line with square markers shows major mergers, and the dash-dotted line with triangle markers shows minor mergers. The increased merger rate with primary galaxy mass results from a higher pair fraction, merger fraction, and shorter merger timescales for massive galaxies. This trend is consistent with hierarchical growth, where high-mass halos experience more rapid accretion and satellite sinking.
}
    \label{rmerge}
\end{figure*}

We calculate the galaxy pair merger fraction in bins of stellar mass, mass ratio, and redshift, defined as
\begin{equation}
C_{\mathrm{merge}} = \frac{N_{\mathrm{merging \, pairs}}}{N_{\mathrm{pairs}}},
\label{eq:cmerge}
\end{equation}
where $N_{\mathrm{merging \, pairs}}$ is the number of galaxy pairs that merge within the same $(\log M_{1}, \mu, z)$ bin, and $N_{\mathrm{pairs}}$ is the total number of pairs in that bin. 

Figure \ref{cmerge} and Figure \ref{histcmerge}  in the Appendix show the resulting merger fractions within mass, mass ratio, and redshift bins. Panel~\ref{cmerge_kate} displays the fractions obtained using our selection criteria (Section \ref{sec:methods}), while panels  \ref{cmerge_puskas}, \ref{cmerge_pfister}, \ref{cmerge_ventou} show the corresponding results when applying the methods of \citet{Puskas2025}, \citet{Pfister2020}, and \citet{Ventou2019}, respectively.  

 In Fig.~\ref{cmerge} we see that in all cases the merger fraction is consistent when using different methods and shows an increase with both stellar mass and redshift. This behaviour is expected: massive galaxies reside in more clustered environments and therefore have a higher probability of encountering and merging with companions. At higher redshift, galaxies have a longer time available to merge successfully by $z=0$, increasing the likelihood that pairs will eventually coalesce. The selections in panel~\ref{cmerge_kate} and ~\ref{cmerge_puskas} show that there is a trade off between $r_{\rm proj}$ and $v_{\rm proj}$ thresholds: excluding pairs at larger separations or higher relative velocities, reduces similarly the number of pairs that ultimately merge.

Figure \ref{tmerge} and Figure \ref{histtmerge} in the Appendix present the distribution of galaxy pair merger timescales. The merger timescale $\mathrm{T_{\rm merge}}$ is defined as the difference between the cosmic age at which the pair is first identified in the lightcone and the cosmic age at which the galaxies merge in the merger tree. For pairs where the two galaxies are detected in different snapshots, we adopt the later snapshot (i.e.~the one closer in time to an observer) as the detection epoch to ensure consistency.  

In Fig.~\ref{tmerge} we show the merger timescales for different pair selection methods. In all four cases (panels \ref{tmerge_kate}, \ref{tmerge_puskas}, \ref{tmerge_pfister}, \ref{tmerge_ventou}), the merger timescale systematically increases toward lower stellar masses. This trend is expected: low-mass galaxies experience weaker dynamical friction, have less concentrated dark-matter halos, and typically reside in lower-density environments, all of which slow the orbital decay of galaxy pairs.

We see a noticeable drop of the timescale for pairs at $z=1.5$, which we explain by the fact that the new refinement level triggered at $z=1.5$ generates sudden higher density, thus speeding up mergers. The higher resolution increases the gas and stellar densities, which can accelerate the final stages of galaxy mergers. In practice, even galaxies that are initially selected at relatively large separations may merge more rapidly once they approach each other at the epoch when the new refinement level is introduced. Therefore, the results around this redshift bin should be taken with caution. This effect affects also the selection of pairs that merge, giving fewer pairs at large separations (see Fig.~\ref{dvdrlim}).

Our optimized selection (panel \ref{tmerge_kate}) yields systematically longer merger timescales at higher redshift than the alternative criteria shown in panels \ref{tmerge_puskas}, \ref{tmerge_pfister}, \ref{tmerge_ventou}. This is a direct consequence of our more extended dynamical limits: by allowing larger values of $\Delta r_{\rm proj}$ and $\Delta v_{\rm proj}$ for massive galaxies, we include pairs that are physically associated but initially more widely separated or with higher orbital energies. These systems still merge, but require additional time to sink toward coalescence. In contrast, the selection schemes in panels~\ref{tmerge_puskas}, \ref{tmerge_pfister}, \ref{tmerge_ventou} adopt stricter or fixed cuts in projected distance and velocity, preferentially selecting tightly bound pairs with shorter sinking times.


\subsection{Galaxy Merger Rate}

To compute the galaxy merger rate, we combine three parameters: the pair fraction $f_{\rm p}$, the merger fraction $C_{\mathrm{merge}}$, and the merger timescale $T_{\mathrm{merge}}$.  

The pair fraction $f_{\rm p}$ is defined as the fraction of galaxies that are in pairs within a given stellar mass, mass ratio, and redshift bin; $C_{\mathrm{merge}}$ is the merger fraction and the merger timescale $T_{\mathrm{merge}}$ represents the average time to coalescence for pairs in the same bin.  

The merger rate is then obtained as

\begin{equation}
 \mathcal{R}_{\mathrm{merge}} = \frac{C_{\mathrm{merge}} \times f_{\rm p}}{T_{\mathrm{merge}}}.
\end{equation}

Figure~\ref{rmerge}  presents the merger rates derived from our sample using the criteria defined in this work. In both cases, observational and simulation-based, the merger rate increases with galaxy mass, illustrating that more massive galaxies experience mergers more frequently.

The resulting merger rates in Fig.~\ref{rmerge} reflect the combined effects of pair fraction, merger fraction, and timescale. The merger rate increases with stellar mass, showing that massive galaxies not only have a higher probability of forming close pairs, but also merge on shorter timescales, similarly to what is found in the Illustris simulation \citep[Figure 11 in][]{Mundy2017}. This mass dependence is consistent with hierarchical growth: high-mass halos experience more frequent accretion events, their satellites sink more rapidly, and their merger-driven growth dominates over in situ star formation at the massive end. 

\begin{figure}
\centering

\includegraphics[width=0.46\textwidth]{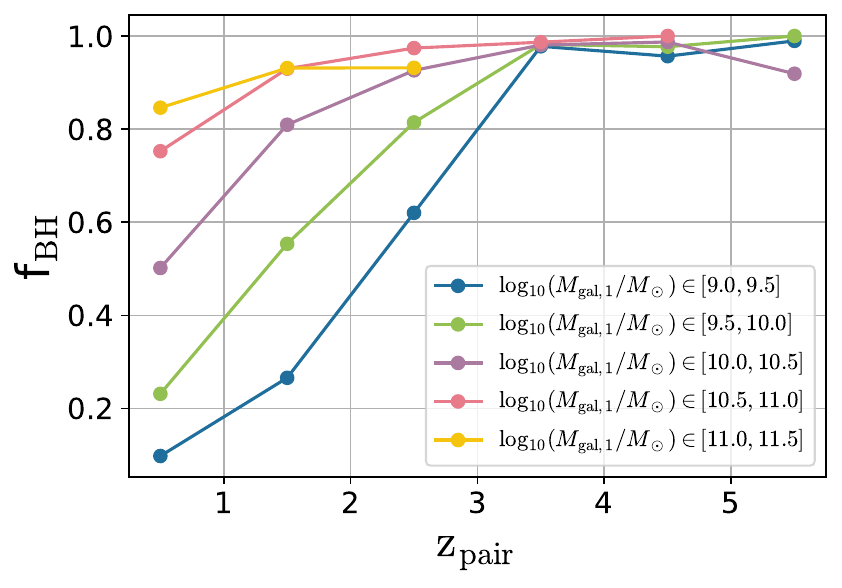}
\caption{Fraction of galaxy pairs hosting at least one BH each, as a function of stellar mass and redshift. Each color represents a different galaxy mass bin. The black hole occupation fraction in galaxy pairs increases with both stellar mass and redshift, consistently with the general occupation fraction \citep{Volonteri2016}.}
    \label{fbh}
\end{figure}

\subsection{Black Hole Mergers and Timescales}

We further investigate the occupation of supermassive  BHs in galaxy pairs that merge. We use the catalog of  BHs from the merger tree, which contains the BH IDs and the host galaxy IDs at each snapshot.  By matching the galaxy IDs and snapshots of our galaxy pairs from the merger tree to the host galaxy IDs and snapshots in the  BH catalog, we identify which galaxies host BHs.

Figure \ref{fbh} and Figure \ref{histfbh} in Appendix show the fraction of merged galaxy pairs where both galaxies in the pair host at least one BH, as a function of stellar mass and redshift. 
These fractions provide an estimate of BH occurrence in the pair population and can help quantify how many merging pairs host at least one BH in each galaxy, which can then be compared to the subsequent  BH merger fractions. Figure~\ref{fbh} shows that the BH occupation fraction in pairs increases  with both stellar mass and redshift, consistently with the general occupation fraction \citep{Volonteri2016}.

\begin{figure*}[tb]
    \centering
    \includegraphics[width=\textwidth]{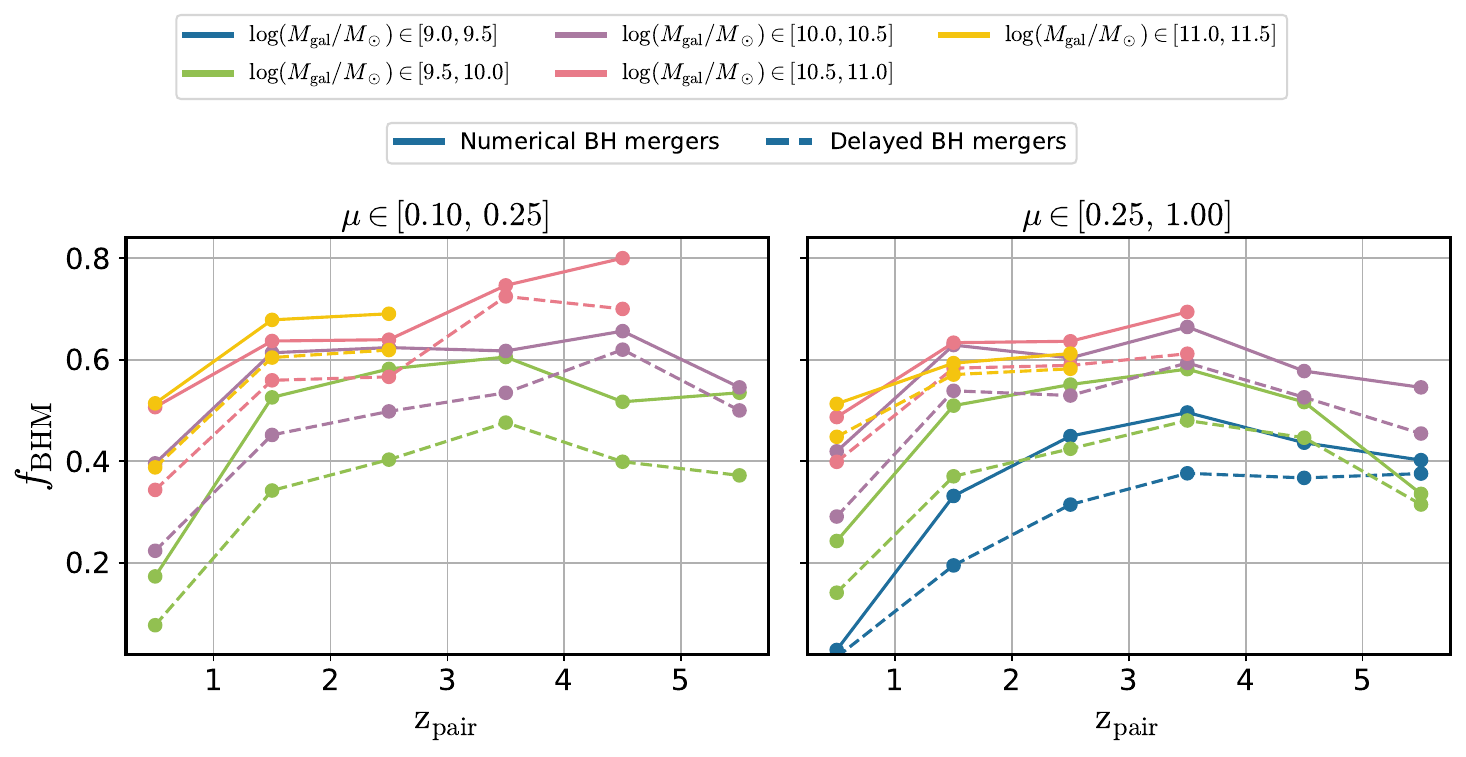}
    \caption{Black hole merger fraction for numerical and delayed mergers. Solid lines represent numerical BH mergers, and dashed lines represent delayed BH mergers.  Each color corresponds to a primary galaxy stellar mass bin. The left panel shows the BH merger fraction for minor galaxy mergers, and the right panel shows the fraction for major galaxy mergers. The BH merger fraction increases with redshift up to $z\sim$3-4, reflecting the limited volume at earlier times, and increases with primary stellar mass, following the trends in black hole occupation and galaxy merger fractions. Numerical mergers yield higher BH merger fractions because the delayed model has fewer BH binaries that actually merge, resulting in a lower overall fraction. 
    }
    \label{fbhm}
\end{figure*}

\begin{figure*}[tb]
    \centering
    \includegraphics[width=\textwidth]{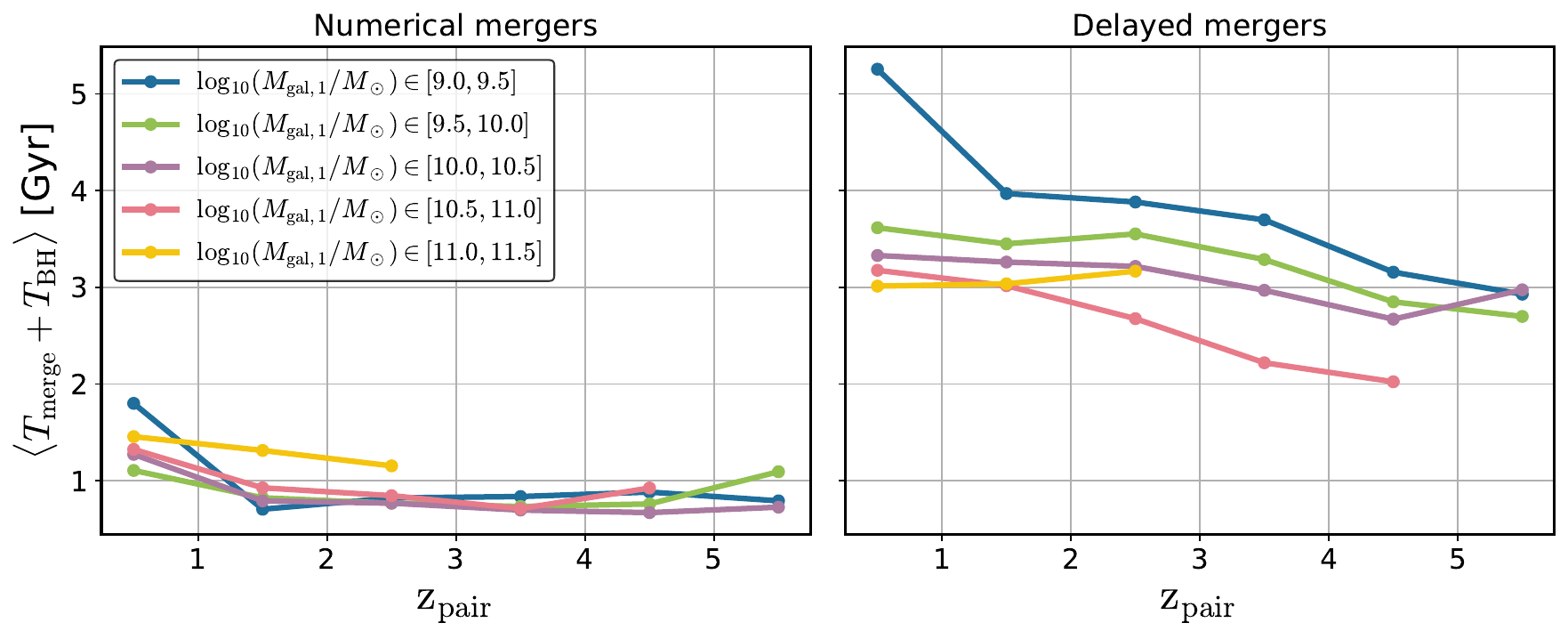}
    \caption{BH merger timescale, calculated as the time between host galaxy pair detection and BH pair merger. Each color represents a different galaxy mass bin. The left panel shows numerical mergers, and the right panel shows delayed mergers. The $x$-axis represents the redshift of the host galaxy pair detection. Numerical timescales show no trend with mass or redshift, because pairs are selected over a wide range of separations. In the delayed model, the timescale clearly decreases with increasing host galaxy mass and redshift, due to the mass dependence in the analytical delay-time model. 
}
    \label{tbhgal}
\end{figure*}

\begin{figure*}[tb]
    \centering
    \begin{minipage}{\textwidth}
        \centering
        \includegraphics[width=\textwidth]{images/legend_mass_and_mu.pdf}
    \end{minipage}

    \vspace{-1cm}  
    \begin{minipage}{\textwidth}
        \centering
        \includegraphics[width=\linewidth]{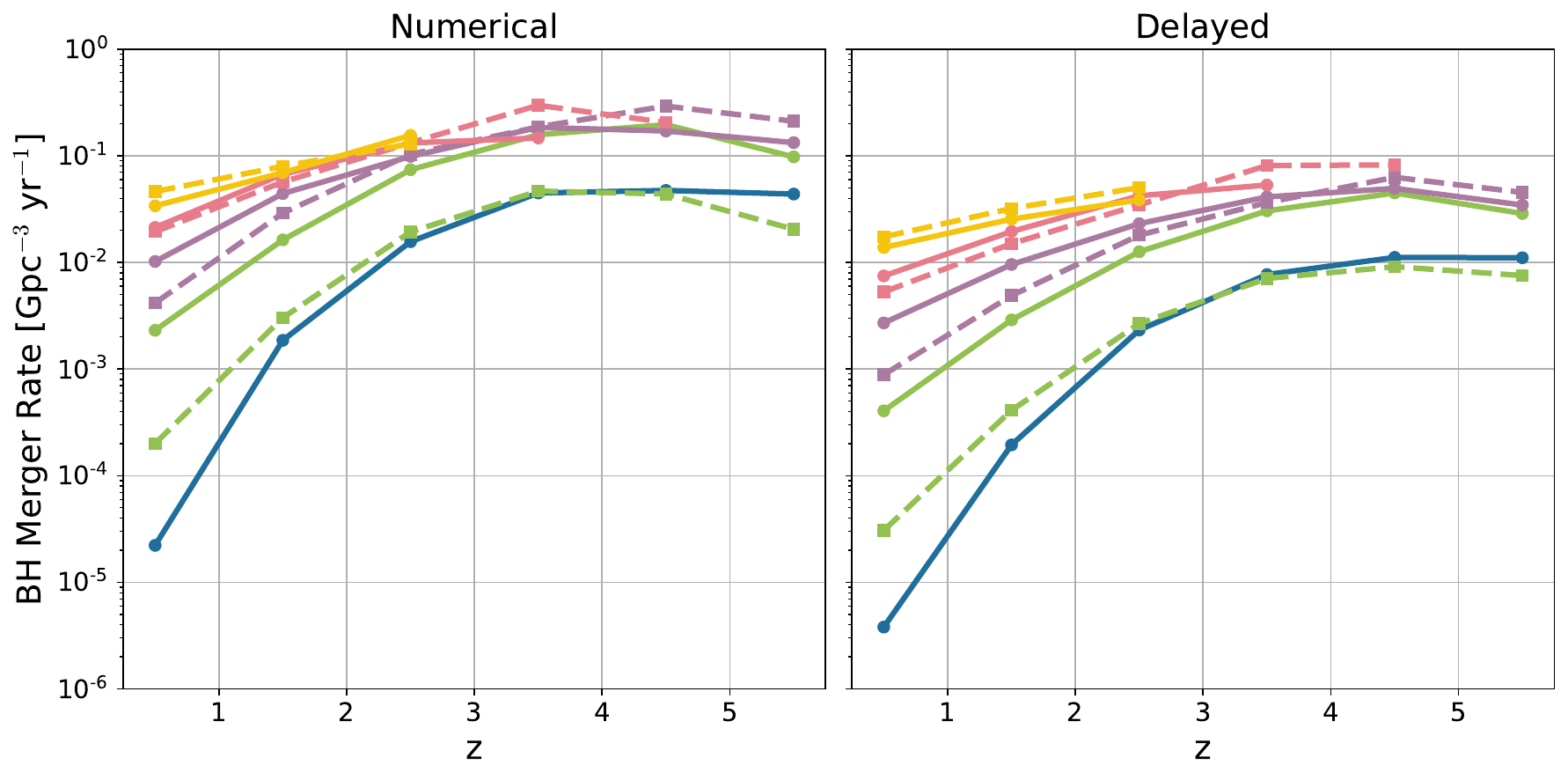}
    \end{minipage}

    \caption{Black hole merger rate, $R_{\mathrm{merge,BH}}$, as a function of galaxy stellar mass and redshift for different galaxy mass ratio bins. The calculation combines the galaxy merger rate with the BH merger fraction and the average BH merger timescale. Solid lines represent major galaxy mergers, and dashed lines represent minor galaxy mergers. The left panel shows the merger rate for numerical BH mergers, and the right panel shows the rate for delayed BH mergers. The black hole merger rate broadly follows the same mass and redshift trends as the galaxy merger rate, but with a lower normalization and a steeper drop toward low redshift due to the additional delay between galaxy and black hole coalescence. The delayed model predicts rates roughly an order of magnitude lower than the numerical model, highlighting how strongly the inferred rate depends on the assumed black hole dynamics.}
    \label{rmergerbh}
\end{figure*}

\begin{figure*}[tb]
    \centering
    \includegraphics[width=\textwidth]{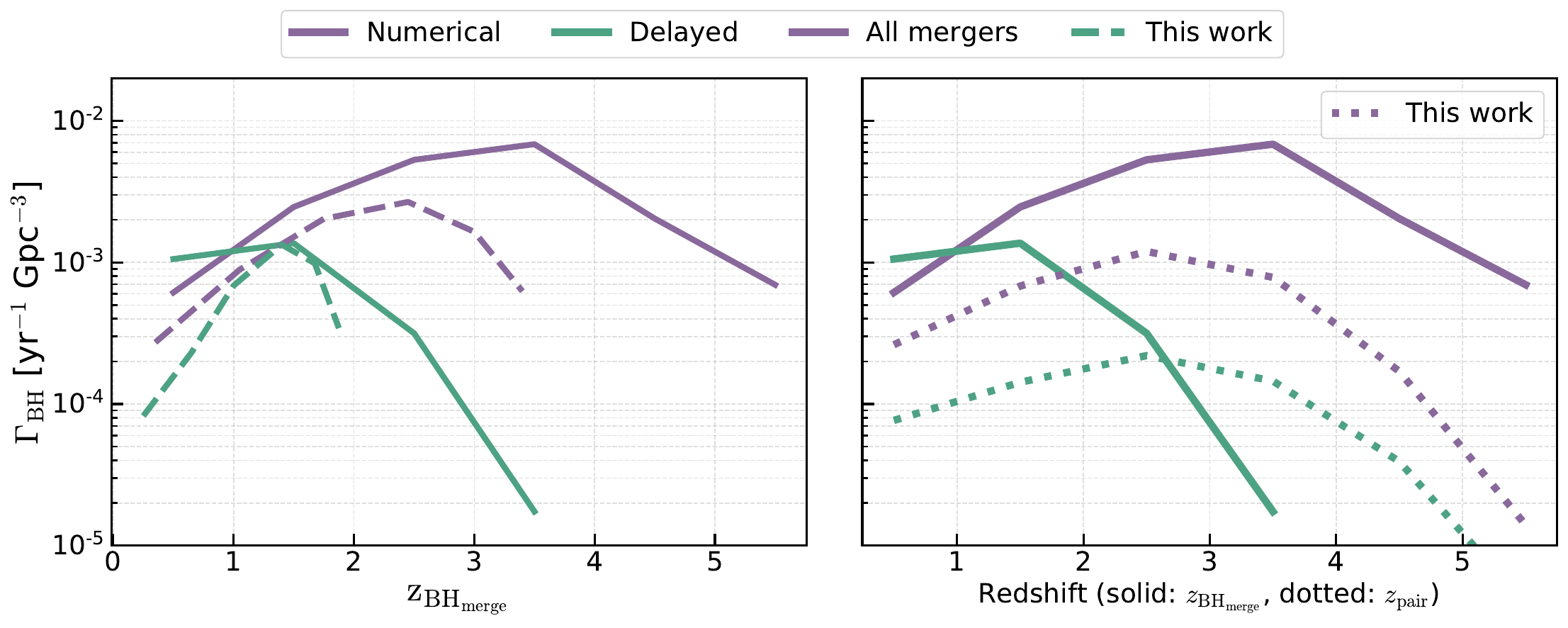}
    \caption{Intrinsic volume-averaged black hole merger rate, $\mathrm{\Gamma_{BH}}$, as a function of cosmic time and redshift. The left panel represents $\Gamma_{\rm BH}(z_{\rm BH_{\rm merge}})$, while the right panel shows $\Gamma_{\rm BH}(z_{\rm pair})$. Solid lines represent the total BH merger rate from the simulation, whereas dashed or dotted lines show the BH merger rate inferred from galaxy pairs. The ``true'' BH merger rate can be obtained from galaxy pair selections by applying correction factors that account for the differences shown here.}
    \label{gamma}
\end{figure*}

\begin{figure*}[tb]
    \centering
    \includegraphics[width=0.8\textwidth]{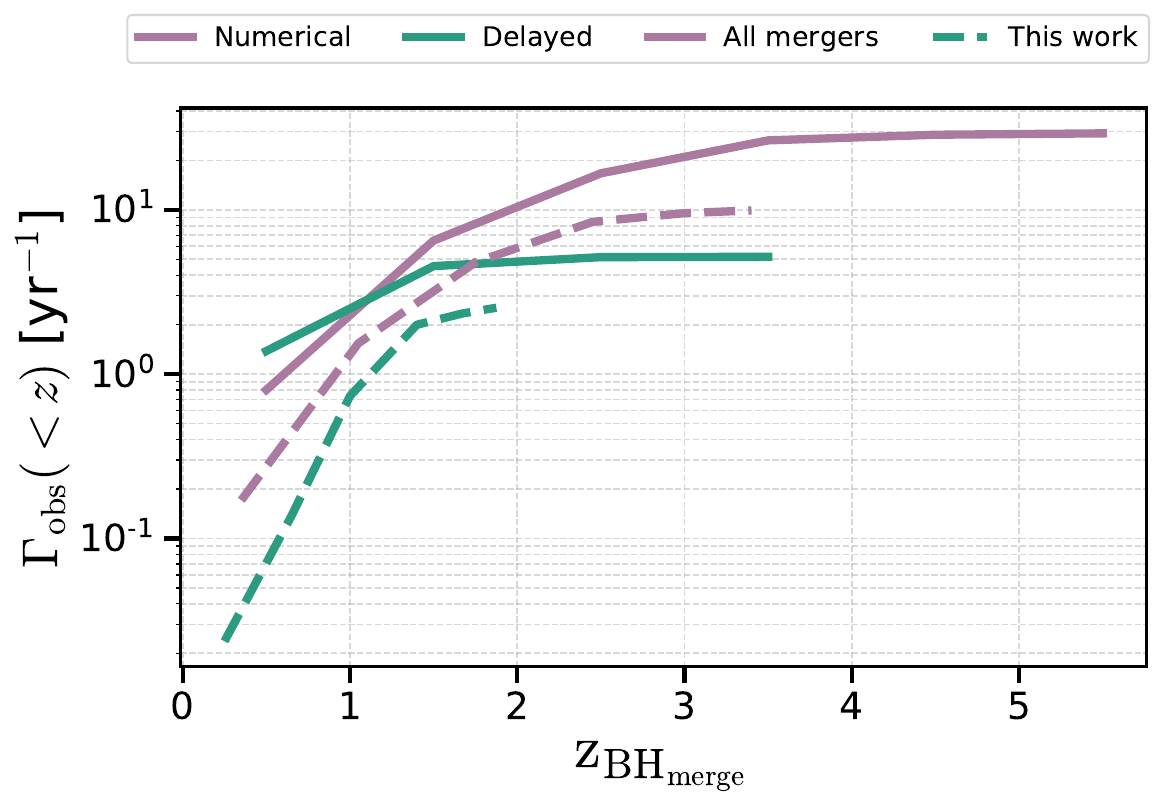}
    \caption{Observable volume-averaged black hole merger rate, $\mathrm{\Gamma_{BH, obs}}$, as a function of redshift.  Solid lines represent the total BH merger rate from the simulation, while dashed lines show BH mergers inferred from galaxy pairs in this work. Purple curves correspond to numerical mergers, and green curves to delayed mergers. The consistent trends and similarities between the total merger rate and the merger rate inferred from galaxy pairs suggests that the latter can be used to estimate the BH merger rate, by applying a correction factor such as the ratio between the solid and dashed curves.}
    \label{gammaobs}
\end{figure*}

We next compute the fraction of  BHs  that merge within galaxy pairs that themselves merge.  We separately investigate numerical and delayed BH mergers, calculating for each bin in stellar mass, mass ratio, and redshift the fraction of BHs that merge.  

To do this, we cross-match the BH merger catalogs with our galaxy pair catalog using the BH IDs and host galaxy IDs at the relevant snapshots.  
A BH merger is counted within a galaxy pair if the following criteria are satisfied:  
\begin{enumerate}
    \item The two BHs must be initially located in the two different galaxies of the merging pair.
    \item Both BHs must appear in the BH merger catalog (numerical or delayed), indicating that they eventually merge.
    \item If a galaxy contains multiple BHs, the criterion is satisfied as long as at least one BH from one galaxy merges with at least one BH from the companion galaxy in the pair.
\end{enumerate}

Figures~\ref{fbhm}, ~\ref{histfbhmnum}, and ~\ref{histfbhmdel} in Appendix present the  BH merger fractions for numerical and delayed mergers, respectively, as a function of stellar mass and redshift. The fraction is defined as the number of galaxy pairs hosting a  BH merger divided by the total number of galaxy pairs within the corresponding redshift, primary stellar mass, and stellar mass ratio bin. The denominator includes all galaxy pairs in the bin, independent of whether they contain no  BHs, a single  BH, or a  BH in each galaxy without undergoing a merger.

In both the numerical and delayed merger models, this fraction increases with redshift up to  $z\sim 3\mbox{--}4$, while at even earlier times the simulation starts missing high-mass galaxies because of its limited volume. 
The BH merger fraction is lower than the galaxy merger fraction: this reflects that even if galaxies merge, the merger of their BHs is not guaranteed. 
The numerical-merger prescription yields systematically higher BH merger fractions than the delayed model. In the numerical case, BHs merge prematurely when they enter the resolution limit, $4 \Delta x$, before a bound binary has physically formed.  In contrast, the delayed prescription enforces a physical inspiral and hardening timescale, preventing artificial early mergers: this causes fewer BH pairs to bind and evolve toward coalescence.  As a result, the delayed model retains a smaller population of BH binaries that ultimately merge, leading to a lower overall BH merger fraction.

Figure~\ref{tbhgal} and Figure~\ref{histtbhgal} in the Appendix show the average BH merger timescale as a function of primary galaxy stellar mass and BH merger redshift, for both the numerical and delayed merger models. 
The  BH merger timescales are plotted as a function of the redshift at which the host galaxy pair was detected. For numerical mergers, the timescale shows no clear dependence on mass or redshift. Trends simply reflect $T_{\rm merge}$, but considering only the subset of galaxies which contain BHs, and whose BHs merge. The longer timescales for BHs in the most massive galaxies reflect the low mass ratios of the merging BHs: mergers between equal mass systems are unlikely for rare massive galaxies, and orbital decay is longer for smaller BHs. The sudden increase in merger timescale for $\log(M_{1}/\msun)\in[9,9.5]$ at the lowest redshift bin is dominated by statistical uncertainty, the bin containing only a few galaxies. 
In contrast, for delayed mergers the timescale decreases with increasing host-galaxy–pair redshift and mass. The mass dependence is clearer in this case because the BH mass, which is tied to the galaxy mass, enters explicitly into the analytical delay-time model \citep[]{Volonteri2020}. We also note that for the highest mass bin there are few galaxies at high redshift, making deriving trends less robust. In the lowest redshift bin  for $\log(M_{1}/\msun)\in[9,9.5]$ there is only one galaxy, therefore the value should be considered fully unconstrained.

\subsection{Black Hole Merger Rates}

We consider three measures of the BH merger rate, each probing a different concept.

\subsubsection{From galaxy mergers to black hole merger rate}

 We start with the definition similar to that of the galaxy merger rate: the number of mergers experienced by a BH per Gyr. It is obtained by linking the galaxy merger rate to the BH occupation fraction and by accounting for the delay between the galaxy merger and the BH merger. 

We compute the  BH  merger rate by rescaling the galaxy merger rate with the fraction of galaxy mergers that lead to BH coalescences and their corresponding merger timescales:
\begin{equation}
\mathcal{R}_{\rm merge,BH}
= \mathcal{R}_{\rm merge}\;
f_{\rm BH}\;
f_{\rm BHM}\;
\left(\frac{T_{\rm merge}}{T_{\rm BH}+T_{\rm merge}}\right),
\label{eq:RmergeBH}
\end{equation}
with all quantities evaluated within bins of mass, mass ratio, and redshift. Here
\begin{align}
f_{\rm BH} &= \frac{N_{\rm pairs\_with\_bh}}{N_{\rm merging\_pairs}}, \\
f_{\rm BHM} &= \frac{N_{\rm BH\_merge}}{N_{\rm pairs\_with\_bh}}.
\end{align}
$N_{\rm pairs\_with\_bh}$ is the number of galaxy pairs where both galaxies host a BH; $N_{\rm merging\_pairs}$ is the number of merging galaxy pairs; $N_{\rm BH\_merge}$ is the number of BH mergers among those pairs. The factor $T_{\rm merge}/(T_{\rm BH}+T_{\rm merge})$ is a correction that accounts for the relative timescales, where $T_{\rm merge}$ is the time from when the galaxy pair is identified until the galaxies merge, and $T_{\rm BH}$ is the time from the galaxy merger to the subsequent BH coalescence.

The total  BH merger rate can therefore be written as
\begin{equation}
\mathcal{R}_{\rm merge,BH} = 
\frac{f_{\rm p} \, C_{\rm merge} \,
f_{\rm BH} \, f_{\rm BHM} }
{T_{\rm BH} + T_{\rm merge}},
\label{eq:RmergeBH_total}
\end{equation}
where $f_{\rm p}$ and $C_{\rm merge}$ are described in equations \ref{eq:fp} and \ref{eq:cmerge}.

Figure~\ref{rmergerbh} shows the BH merger rate as a function of stellar mass and redshift for different merger mass ratios. The BH merger rate follows the same general trends as the galaxy merger rate, increasing with stellar mass and redshift, but with a lower normalization. The BH merger rate drops at lower redshifts; this behaviour reflects the delay between the galaxy merger and the subsequent BH coalescence. Although galaxy mergers are more frequent at high redshift, many BH mergers occur later, once sufficient time has elapsed for coalescence, and the BHs of low-redshift pairs may never have a chance to merge before $z=0$.

\subsubsection{Volume-averaged black hole merger rate density}

We compute the intrinsic volume-averaged BH merger rate density as
\begin{equation}
\Gamma_{\rm BH}(z) \;=\; \sum_{M_{\rm gal,1}} \mathcal{R}_{\rm merge,BH} (M_{\rm gal,1},z) \Phi_{gal}(M_{\rm gal,1},z)
\label{eq:GammaBH}
\end{equation}
where $\Phi_{gal}(M_{\rm gal,1},z)$ is the galaxy mass function within the chosen mass and redshift bin derived from the simulation \citep{Kaviraj2017}. 
Calculating $\Gamma_{\rm BH}(z_{\rm pair})$ at the redshift of the galaxy pair selection is straightforward, while obtaining $\Gamma_{\rm BH}(z_{\rm BH_{merge}})$ at the time when the BHs merge requires shifting the elements of each bin from $z_{\rm pair}$ to $z_{\rm BH_{merge}}$ via $T_{merge}$.  

In Figure~\ref{gamma}, solid lines indicate the total BH merger rate density from the full simulation volume, whereas dashed lines show the contribution obtained from BH mergers associated with our selected merging galaxy pairs.

$\Gamma_{\rm BH}(z_{\rm BH_{merge}})$ inferred from galaxy pairs  closely follows the true BH merger rate in the simulation in the case of numerical mergers. For delayed mergers, galaxy pair selection leads to an underestimate the BH merger rate at low redshift ($z\lesssim 1$) as low redshift galaxy pairs have less time for BH mergers to complete before $z=0$ (cf. the discussion on $C_{\rm merge}$). Similar conclusion can be drawn on $\Gamma_{\rm BH}(z_{\rm pair})$. 
This shows that galaxy pair selections can be converted into a BH merger rate by applying a correction factor, given by the ratio between the solid and dashed or dotted curves.

\subsubsection{Observable black hole merger rate}

The intrinsic merger rate density of compact objects can be converted to the \emph{observable} all-sky merger rate (per observer year) through integration over comoving volume and redshift:

\begin{equation}
\Gamma_{{\rm BH},{\rm obs}}
\;=\;
\int_{0}^{z_{\max}}
\Gamma_{\rm BH}(z)\;
\frac{\mathrm{d}V_{\rm c}}{\mathrm{d}z}\;
\frac{1}{1+z}\;
P_{\rm det}(z,\theta)\;\mathrm{d}z,
\label{eq:GammaBHobs}
\end{equation}
where $P_{\rm det}(z,\theta)$ is the detection probability for sources with parameters $\theta$ (e.g.\ masses, spins, sky position, orientation), and the all-sky comoving volume element is
\begin{equation}
\frac{\mathrm{d}V_{\rm c}}{\mathrm{d}z} \;=\; 4\pi \;\frac{c\,d_{\rm c}^2(z)}{H(z)}.
\end{equation}
Setting $P_{\rm det}=1$ yields the total number of BH mergers occurring per observer year across the whole sky, irrespective of detectability. Here, $d_c$ denotes the comoving distance to the source, while $H(z)$ is the
redshift-dependent Hubble parameter describing the expansion rate of the Universe at redshift $z$.

Figure~\ref{gammaobs} presents the observed volume-averaged BH merger rate, $\mathrm{\Gamma_{BH, obs}}$. For numerical mergers, the observable rate is dominated by BHs at $z \gtrsim 2$, in contrast, the delayed merger model predicts that the main contribution comes from lower redshift systems. This difference arises because the delayed prescription suppresses premature coalescences of low-mass BHs and gives shorter timescales for more massive BHs, furthermore it allows massive BH binaries to survive longer before merging, shifting their contribution to later stages where their masses are higher. We find a similar relationship between the BH merger rate inferred from galaxy pair selection and the full merger rate as for $\Gamma_{\rm BH}$ (Fig.~\ref{gamma}), indicating that the former can be converted into the latter by applying a correction factor in an analogous way.

\clearpage

\section{Conclusions}
\label{sec:conclusion}

In this paper we have tested the ability of galaxy pair selection to be converted into a BH merger rate, by analyzing galaxy pairs and BHs via the Horizon-AGN simulation and its lightcone. 
We developed mass, mass ratio, and redshift dependent dynamical criteria for selecting merging galaxy pairs, optimized for purity and completeness. We then analyzed the galaxy and BH merger rate and their connection, when both are derived from pair counts.

\begin{itemize}

    \item {We compared our optimized selection to several commonly used criteria from
    both observational-based studies
    \citep{Puskas2025, Ventou2019, Conselice2022}. These comparisons show that adaptive limits can recover the expected physical trends with mass and redshift more robustly than fixed projected-distance or velocity cuts, as also shown by \citet{OLeary2021}}.

    \item {Across all selection methods, the galaxy merger fraction increases with stellar mass and redshift. This trend reflects the more clustered environments of massive galaxies and the shorter dynamical times at high redshift. Differences between selection schemes arise primarily from variations in the allowed $ \Delta r_{\rm proj}$ and $\Delta v_{\rm proj}$ ranges.} 

    \item  The timescales derived from our optimized selection are generally longer than those from more restrictive criteria, reflecting our less conservative limits, but our results demonstrate that dynamically motivated, mass and redshift-dependent selection limits can yield higher galaxy merger rates.

    \item    We connected BHs in galaxy pairs to BHs merging in the simulation, obtaining a BH merger timescale that can be added to the galaxy merger timescale. The numerical and delayed merger models produce qualitatively different timescale and mass dependences that reflect the need for clarity in BH dynamics in simulations.

\item We show how the BH merger rate derived from galaxy pairs can be converted into a total BH merger rate, highlighting that galaxy pair counts can be a powerful tool to predict BH mergers once correction factors are applied. 

    \item    The resulting BH merger rates show dependence on redshift and merger timescales, which in turn depend on galaxy and BH masses. These differences highlight the sensitivity of BH merger rate predictions to the adopted merger model.\\

\end{itemize}

Overall, our results demonstrate that connecting galaxy mergers to BH coalescences requires both physically motivated galaxy-pair selection and a realistic treatment of unresolved BH dynamics in simulations. The choice of BH merger model has a significant impact on the inferred merger timescales and rate densities, with important implications for predictions of gravitational-wave event rates. We note, however, that once progress is obtained in estimating the BH merger timescales from simulations  \citep{Tremmel2015,Chen2022,Mannerkoski2023,Li2025,Holley2025}, the BH merger rate inferred from galaxy pairs can be easily converted in a total BH merger rate, as the two trace each other very well.

\section{Acknowledgements}

EL is grateful for financial support from the Nedelandse Organisatie voor Wetenschappelijk Onderzoek (NWO) through the NWO/Groot grant  "Gravitational waves: The new cosmic messengers".

This work has received funding from the Swiss State Secretariat for Education, Research and Innovation (SERI) under contract number MB22.00072, as well as from the Swiss National Science Foundation (SNSF) through project grant 200020\_207349.

The Cosmic Dawn Center (DAWN) is funded by the Danish National Research Foundation under grant DNRF140.

This work was made possible by funding from the French National Research Agency (grant ANR-21-CE31-0026, project MBH\_waves) and from the Centre National d’Etudes Spatiales (MV).

\section{Data Availability}

The Horizon-AGN simulation data used in this work are publicly available at \url{http://www.horizon-simulation.org}. Derived data products and analysis scripts can be shared upon reasonable request to the corresponding author.

\bibliographystyle{mnras}
\bibliography{mergers} 

\clearpage
\appendix
\captionsetup{justification=centering} 

\begin{figure*}[bh]
    \centering
    \includegraphics[width=\textwidth]{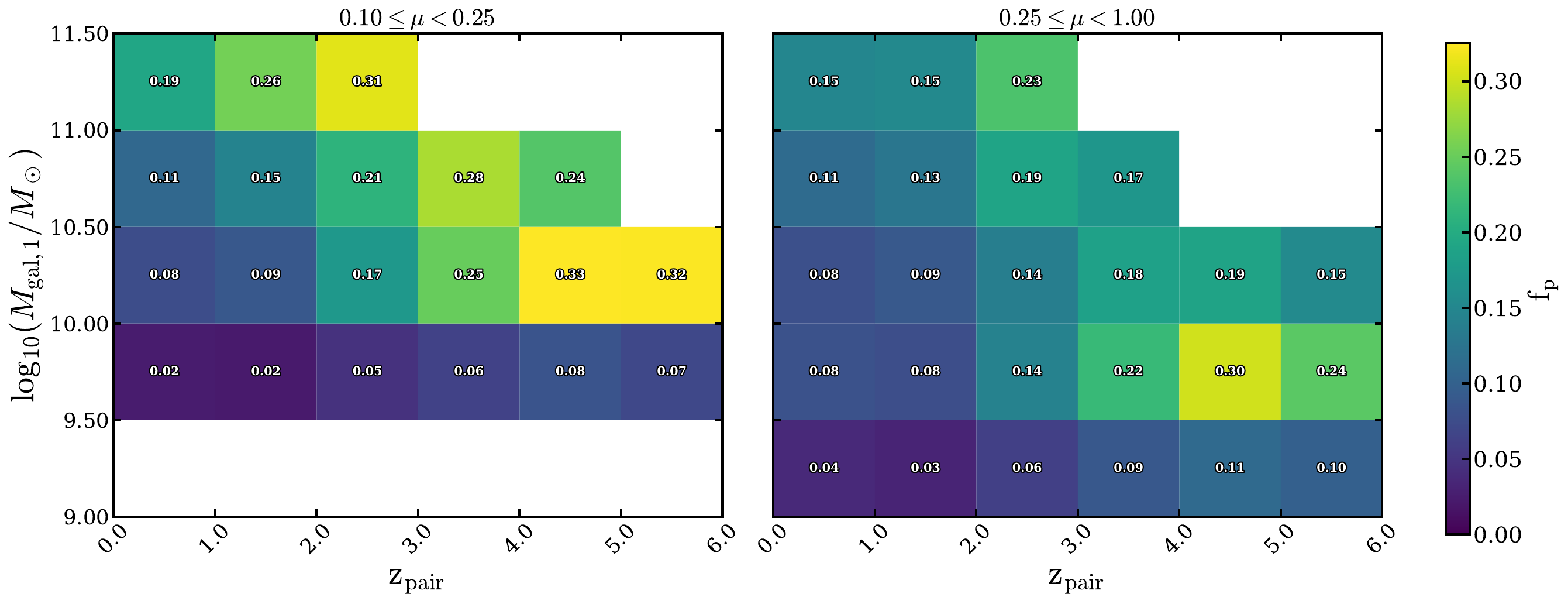}
    \caption{2D histogram of the galaxy pair fraction for each stellar mass and redshift bin, with the left panel showing minor and the right panel showing major galaxy mergers.}
    \label{histfp}
\end{figure*}

\begin{figure*}[tb]
    \centering
    \includegraphics[width=\textwidth]{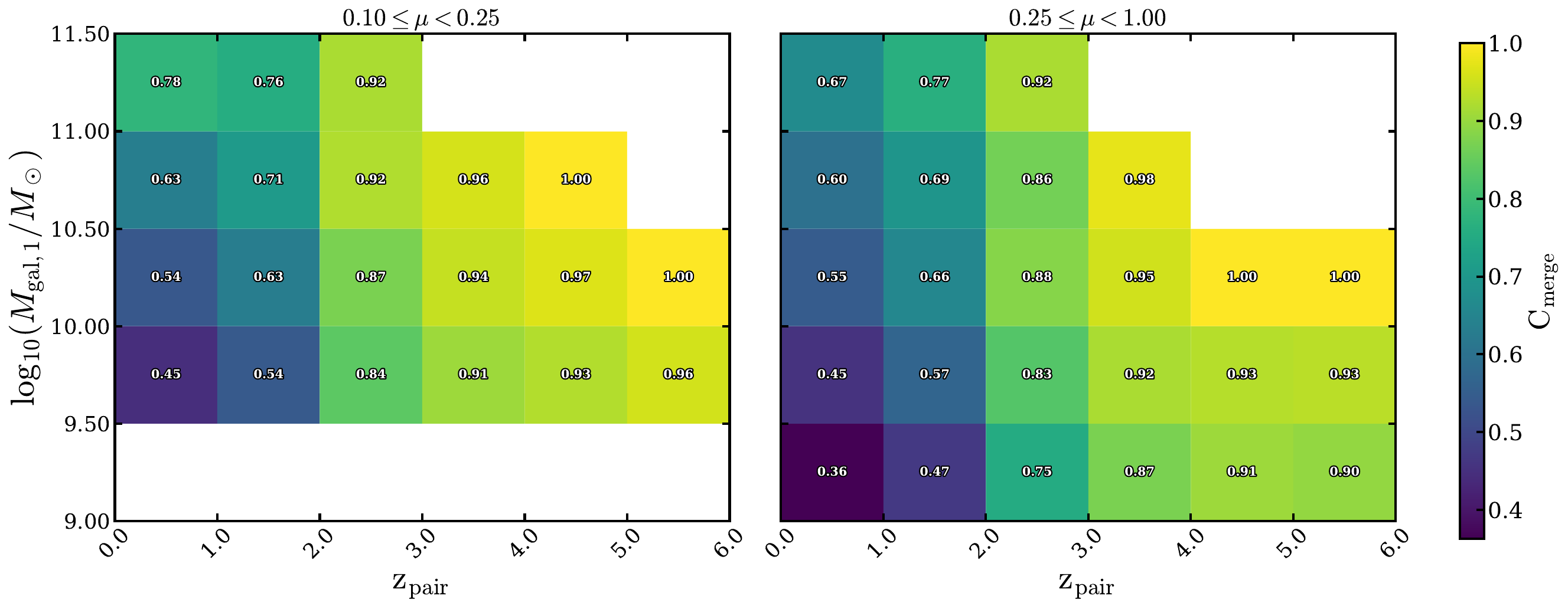}
    \caption{2D histogram of the galaxy merger fraction for each stellar mass and redshift bin, with the left panel showing minor and the right panel showing major galaxy mergers.}
    \label{histcmerge}
\end{figure*}

\begin{figure*}[tb]
    \centering
    \includegraphics[width=\textwidth]{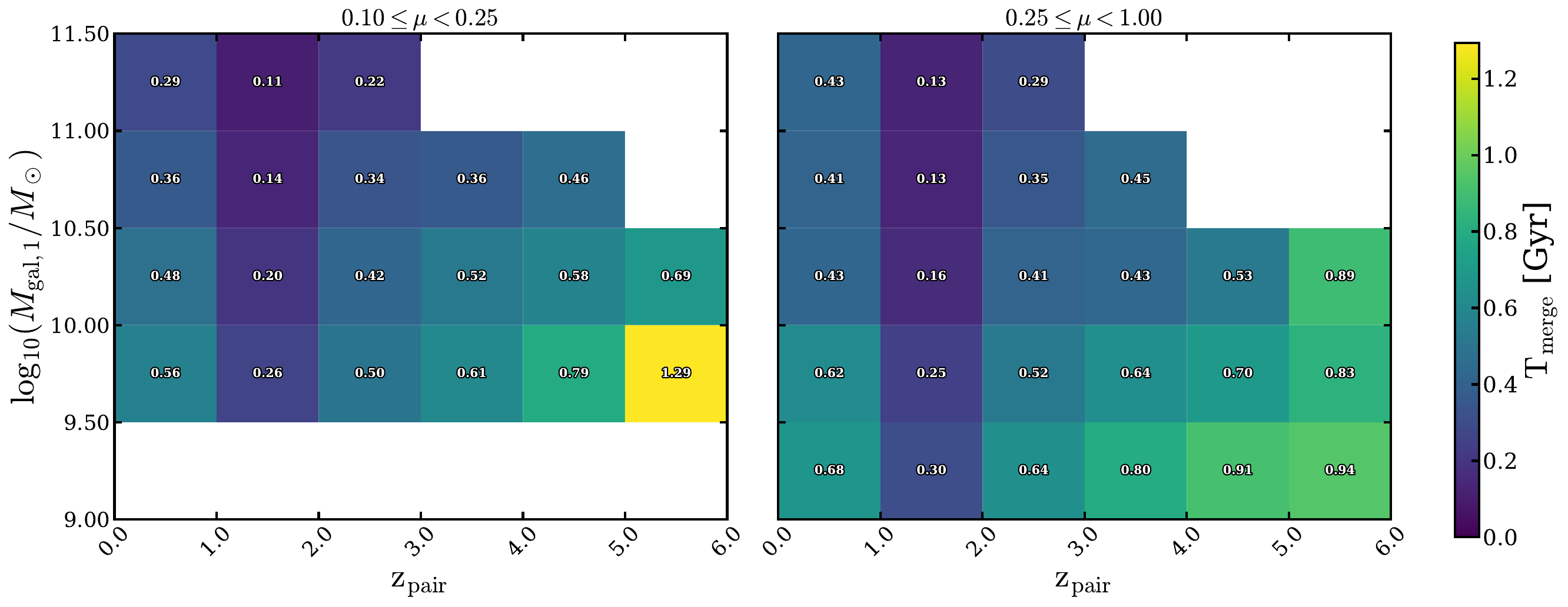}
    \caption{2D histogram of the average galaxy merger timescale for each stellar mass and redshift bin, with the left panel showing minor and the right panel showing major galaxy mergers.}
    \label{histtmerge}
\end{figure*}

\begin{figure*}[tb]
    \centering
    \includegraphics[width=\textwidth]{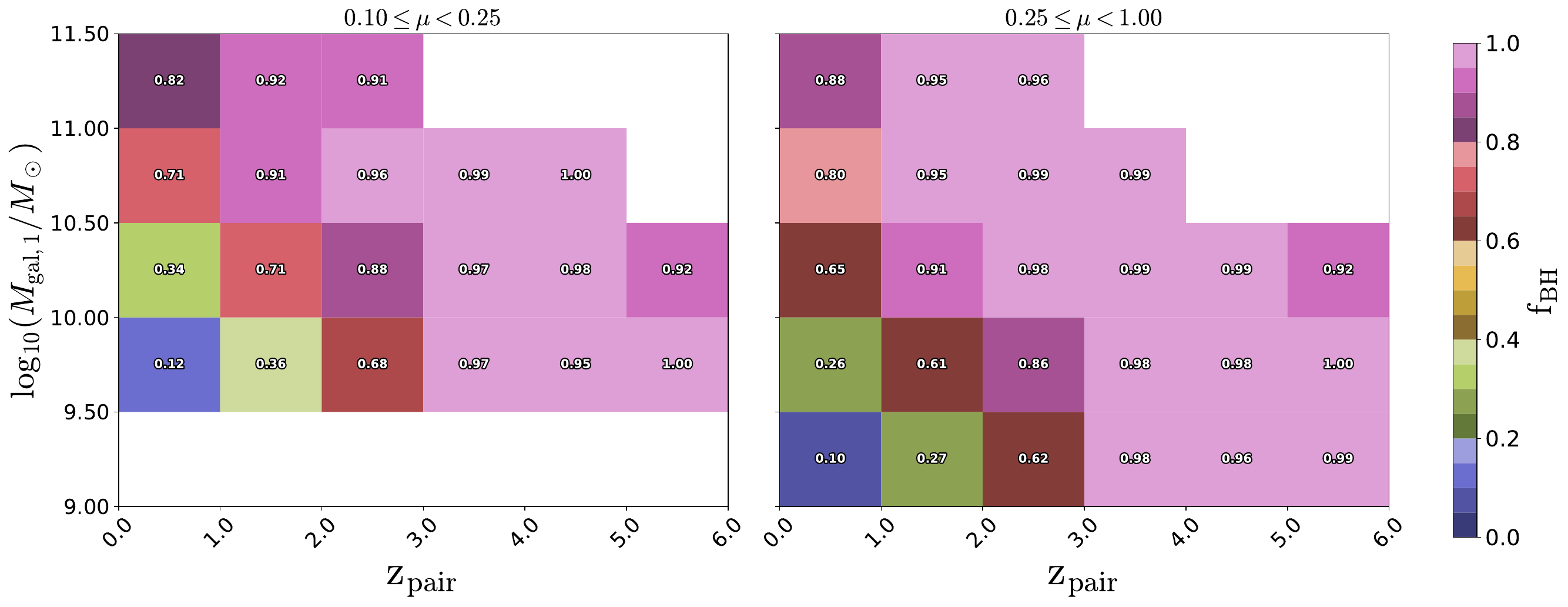}
    \caption{2D histogram of the fraction of galaxy pairs where both galaxies host at least one BH ($f_\mathrm{p}$) for each galaxy stellar mass and redshift bin, with the left panel showing minor and the right panel showing major galaxy mergers.}
    \label{histfbh}
\end{figure*}

\begin{figure*}[tb]
    \centering
    \includegraphics[width=\textwidth]{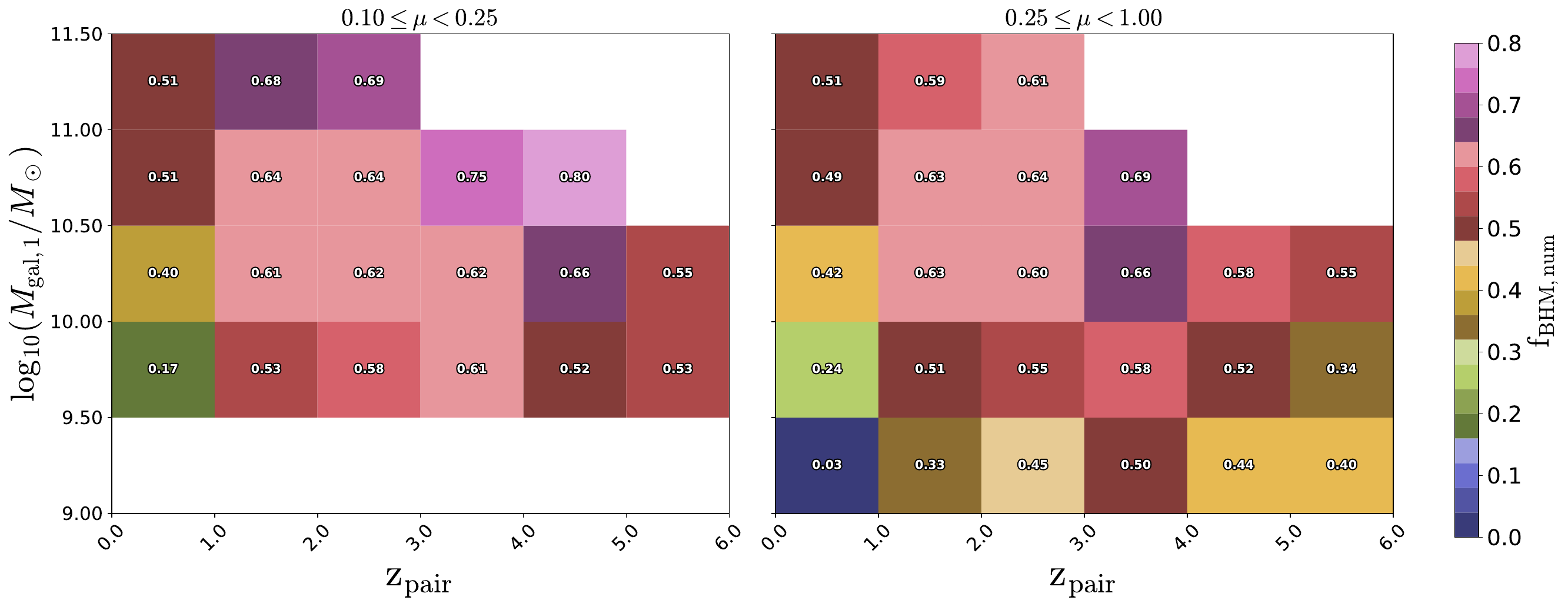}
    \caption{2D histogram of the BH merger fraction for numerical mergers for each galaxy stellar mass and redshift bin, with the left panel showing minor and the right panel showing major galaxy mergers.}
    \label{histfbhmnum}
\end{figure*}

\begin{figure*}[tb]
    \centering
    \includegraphics[width=\textwidth]{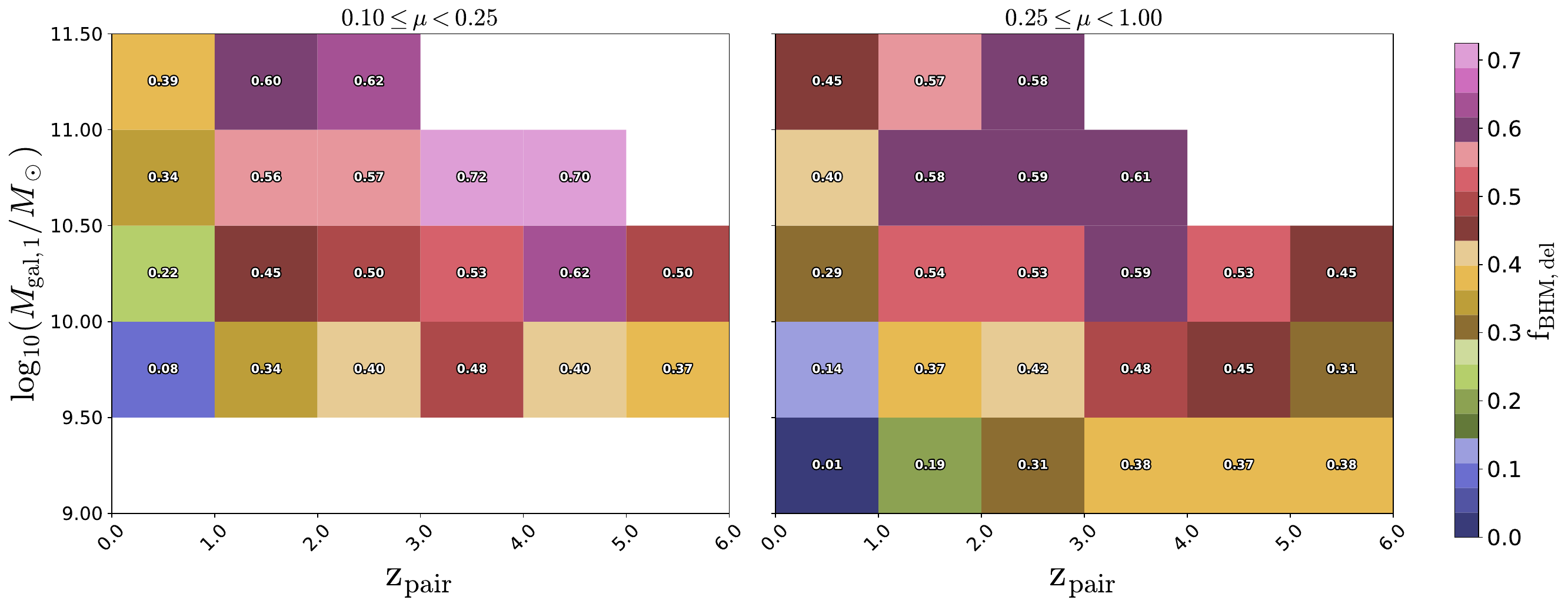}
    \caption{2D histogram of the BH merger fraction for delayed mergers for each galaxy stellar mass and redshift bin, with the left panel showing minor and the right panel showing major galaxy mergers.}
    \label{histfbhmdel}
\end{figure*}

\begin{figure*}[tb]
    \centering
    \includegraphics[width=\textwidth]{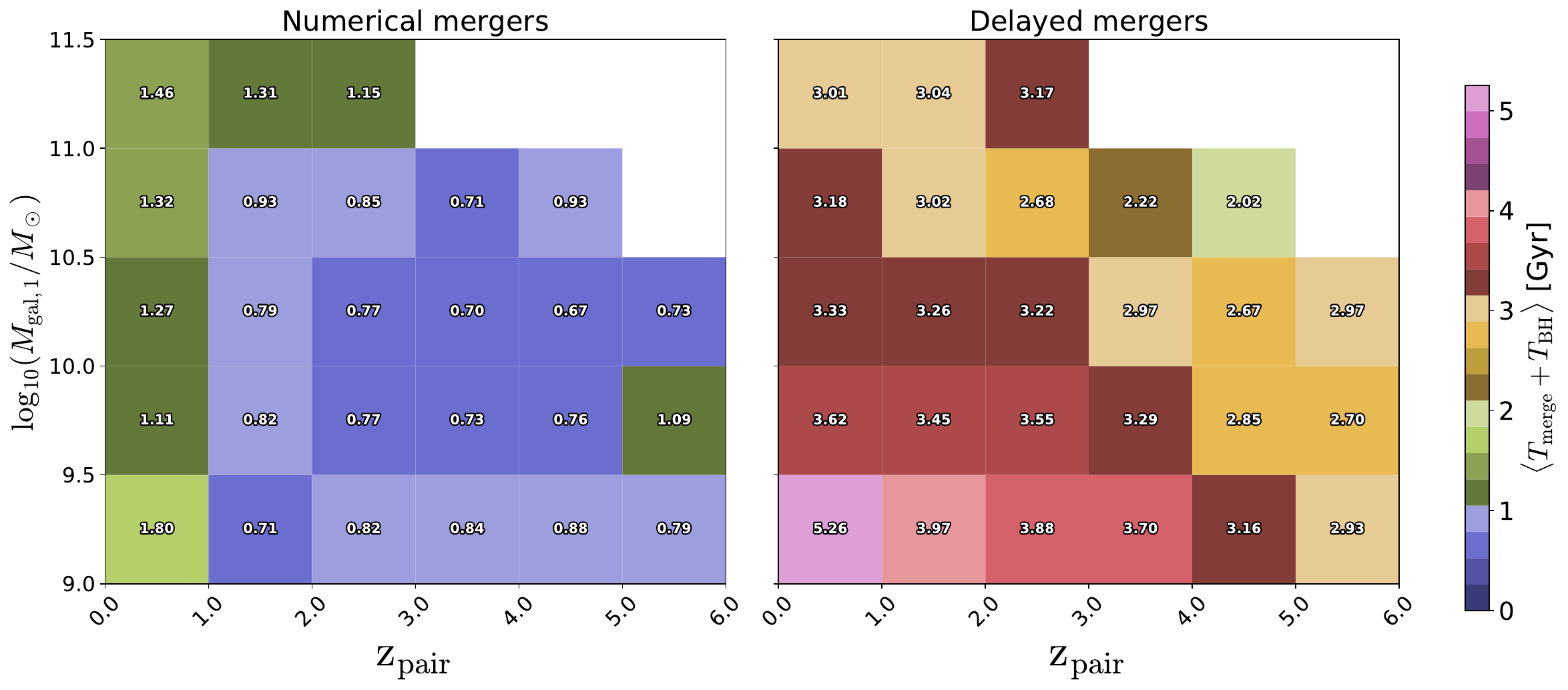}
    \caption{2D histogram of the black hole merger timescale in each galaxy mass and BH merger redshift bin, with the left panel showing numerical BH mergers and the right panel showing delayed BH mergers.}
    \label{histtbhgal}
\end{figure*}

\end{document}